\documentclass{article}
\PassOptionsToPackage{numbers, compress}{natbib}

    \usepackage[preprint]{neurips_2022}

\usepackage{color,array}
\usepackage{graphicx}
\usepackage{amsmath}
\usepackage{amssymb}
\usepackage{caption}
\usepackage{balance}
\usepackage{listings}
\usepackage{color}
\usepackage{float}
\usepackage[utf8]{inputenc} 
\usepackage[T1]{fontenc}    
\usepackage{hyperref}       
\usepackage{url}            
\usepackage{booktabs}       
\usepackage{amsfonts}       
\usepackage{nicefrac}       
\usepackage{microtype}      
\usepackage{lipsum}		 
\usepackage{graphicx}
\usepackage{natbib}
\usepackage{doi}
\usepackage{multirow}
\usepackage{mathbbol}
\usepackage{algorithm}
\usepackage{algpseudocode}
\usepackage[utf8]{inputenc} 
\usepackage[T1]{fontenc}    
\usepackage{hyperref}       
\usepackage{url}            
\usepackage{booktabs}       
\usepackage{amsfonts}       
\usepackage{nicefrac}       
\usepackage{microtype}      
\usepackage{xcolor}         
\usepackage[edges]{forest}
\usetikzlibrary{shadows.blur}

\title{Deep Neural Network Techniques for Monaural Speech Enhancement and Separation: State of the art Analysis}

\author{%
  Peter Ochieng \\
  Department of Computer Science.\\
  University of Cambridge.\\
   \texttt{po304@cam.ac.uk.} \\
}

\begin{document}

\maketitle

\begin{abstract}
 Deep neural networks (DNN) techniques have become pervasive in domains such as natural language processing and computer vision. They have achieved great success in tasks such as machine translation and image generation. Due to their success, these data driven techniques have been applied in audio domain. More specifically, DNN models have been applied in speech enhancement and separation to achieve denoising, dereverberation, speaker extraction and speaker separation in monaural speech intelligibility improvement. In this paper, we review some dominant DNN techniques being employed to achieve speech enhancement and separation. The review looks at the pipeline of speech enhancement and separation techniques  from feature extraction, how DNN-based tools models both global and local features of speech and model training (supervised and unsupervised). The review also covers the use of domain adaptation techniques and  pre-trained models to boost speech enhancement process.

\end{abstract}

\section{Introduction}
Techniques for monaural speech intelligibility improvement can be categorised  either as speech enhancement or separation. Speech enhancement involves isolating a target speech either from noise \cite{Bando2018} or a mixed speech \cite{xiao2019single}. Speech enhancement involves tasks such as dereverberation, denoising and speaker extraction. Speaker separation on  the other hand seeks to estimate independent speeches composed in a mixed speech \cite{Wang2013}. Speech enhancement and separation have applications in multiple domains such as automatic speech recognition, mobile speech communication and designing of hearing aids  \cite{Wang20142}.  Initial research on speech enhancement and separation exploited techniques such as non-negative matrix factorization \cite{Schmidt2006} \cite{Wang201422} \cite{Virtanen2009} probabilistic models  \cite{Virtanen2006} and computational auditory scene analysis (CASA)\cite{Shao2006}. However, these techniques are tailored for  closed-set speakers (i.e., do not work well with mixtures with unknown speakers) which significantly restricts their applicability in  real environments. Due to the recent success of deep learning models in different domains such natural language processing and computer vision, these data driven techniques have been introduced to process audio dataset. In particular,  DNN models have become popular in speech enhancement and separation and have achieved great performance in terms of boosting speech intelligibility and their ability to enhance speech with unknown speakers \cite{Luo2019} \cite{Subakan2021}. In order to be effective in speech enhancement and separation, DNN models must extract important features of speech, maintain order of audio frames, exploit both local and global contextual information to achieve coherent separation of speech data. This necessitates that DNN models should include techniques tailored to meet these requirements. Discussion of these  techniques is the core subject of this review.  Further, in computer vision and text domain, large pre-trained models are used to extract universal representations that are beneficial to downstream tasks.  The review discusses the impact of pre-trained models to the speech enhancement and separation domain. It also discusses DNN techniques being adopted by speech enhancement and separation tools to reduce computation complexity to enable them work in low latency and resource constrained environments. The review therefore focuses on the whole pipeline of DNN application to speech enhancement and separation, i.e., from feature extraction, model implementation, training and evaluation. Our goal is to uncover the dominant techniques at each level of DNN implementation. In each section, we highlight key emerging features and challenges that exist. A recent review \cite{Wang2022} only looked at supervised techniques of performing speech separation and in this review, we discuss both supervised and unsupervised methods. Moreover, with the fast-growing field of deep learning, new techniques have emerged that necessitates a new look into how these techniques have been implemented in speech enhancement and separation. The review is constrained to discussing how DNN techniques are being applied to monaural speech enhancement and so we do not focus on multi-channel speech separation (which has been covered in \cite{Gannot2017}). 
\par The paper first explains the types of speech enhancement and separation (section 2) by highlighting their key elements  and the tools that focus on each type. It discusses the key speech features that are being used by speech enhancement and separation tools in section 3. This section looks at how the features are derived and how they are used to train the DNN models in supervised learning technique. Section 5 discusses the techniques the tools use to model long dependencies that exist in speech. The paper discusses model size compression techniques in section 6. In section 7, the paper discusses some of the popular objective functions used in speech enhancement and separation. Section 8 discusses how some tools are implementing unsupervised techniques to achieve speech enhancement and separation. Section 9 discusses how the speech separation and enhancement tools are being adapted to the target environment. In section 10 the paper looks at how pre-trained models are being utilized in the speech enhancement and separation pipeline. Finally, section 11 looks at future direction. Figure 1 gives an overall organization and topics covered by the paper.
\begin{figure} 
  \centering  
\scalebox{0.51}{
\begin{forest}
  forked edges,
  for tree={
    edge+={thick},
    inner color=gray!5,
    outer color=gray!20,
    rounded corners=2pt,
    draw,
    thick,
    tier/.option=level,
    align=center,
    font=\sffamily,
    blur shadow,
  },
  where level<=1{}{
    rotate=90,
    anchor=east,
  },
  [Deep Neural Network Techniques for Monaural Speech Enhancement and Separation
    [, coordinate, 
      [Types of SE and separation
      [Speech separation]
      [Dereverberation]
      [Denoising]
      [Speech extraction]]]
      [Features
        [Fourier spectrum  [Log-power spectrum features][Mel-frequency spectrum features][DFT magnitude features][Complex DFT features][Complementary features][Supervised Training [Spectral Mapping Techniques][Spectral Masking Techniques][Generative Modelling]] [Phase handling]]
        [Time domain [Supervised training[Adaptive front-end method][Waveform mapping][Generative modelling]]]
      ]
      [Long term\\ dependency modelling [RNN] [TCN] [Use of transformers]]
      [Model size\\ reduction techniques, calign with current
        [Use of dilated convolution]
        [Parameter quantization] [Use of depthwise separable convolution] [Knowledge distillation] [Parameter pruning] [Weight sharing:]
      ]
      [Objective functions
        [SSTOI]
        [SI-SDR, calign with current]
        [PMSQR]
      ]
       [Unsupervised \\Techniques]
        [Domain\\ Adaptation\\ Techniques]
         [Use of \\Pre-trained \\Models]
         [Future\\ directions]]
    ]
    [expatriate, inner color=green!10, outer color=green!25, child anchor=west, edge path'={(!u.parent anchor) -| (.child anchor)}, before drawing tree={y'+=7.5pt} ]
  ]
\end{forest}}
\caption{Overall structure of the topics covered by this review.}\label{Overall structure of the topics covered by this review}
\end{figure}
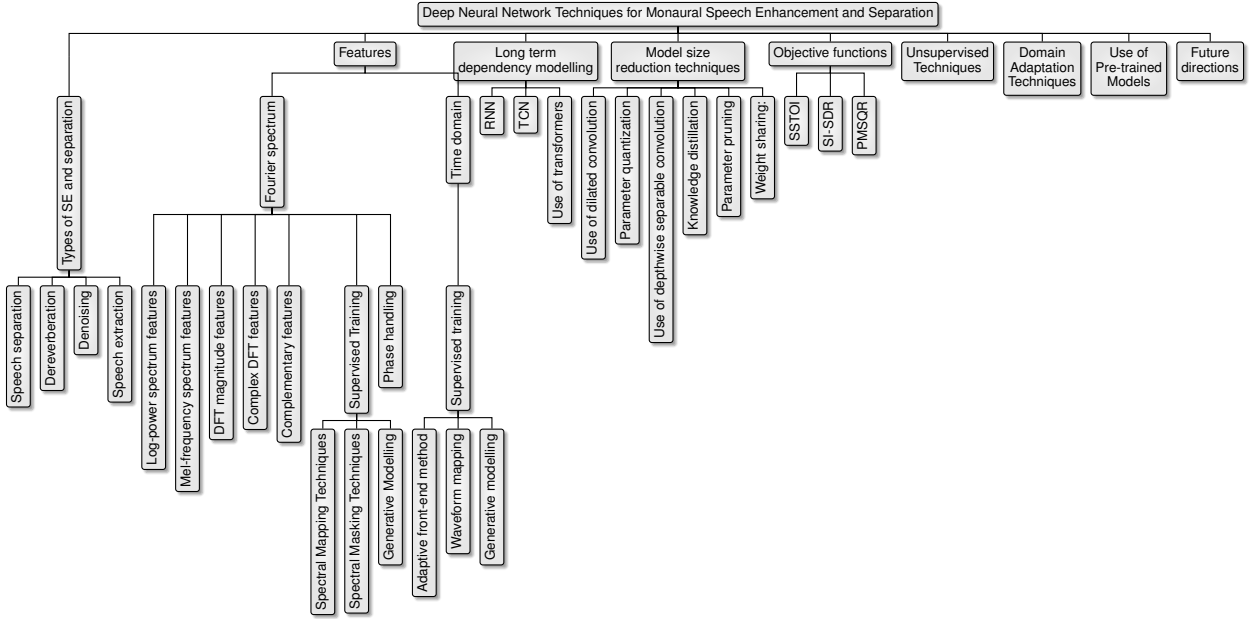
\section{Types of speech separation and enhancement}
\subsection{Speech separation}
Scenarios arise where more than one target speech signals are  composed in a given speech mixture and the goal is to isolate each independent speech composed in a mixture. This problem is known as  speech  separation. For a mixture that is composed of $C$ independent speech signals $x_c(n)$ with $c=1,\cdots,C$, a recording $y(n)$ composed of the $C$ speech signals can be represented as:
\begin{equation}
    y(n)=\sum_{c=1}^{C}x_c(n)
\end{equation}
Here, $n$ indexes time. The goal of speech separation is to estimate each independent $x_c$ speech signal  composed in $y(n)$. Separating speech from another speech is a daunting task by the virtue that all speakers belong to the same class and share similar characteristics\cite{Hershey2016}. Some models such as \cite{Wang2017} and \cite{Wang2017b} lessen this  by performing speech separation on a mixed speech signal based on gender voices present. They exploit the fact that there is large discrepancy between male and female voices in terms of vocal track, fundamental frequency contour, timing, rhythm, dynamic range etc. This results in a large spectral distance between male and female speakers in most cases to facilitate a good gender segregation. For speech separation that 
  the mixture  involves speakers of the same gender, the separation task is much difficult since the pitch of the voice  is in the same range \cite{Hershey2016}. Most  speech  separation tools that solve this  task  such as \cite{Zeghidour2021} \cite{Conference2014} \cite{Weng2015} \cite{Isik2016} \cite{Hershey2016} and  \cite{Luo2019} cast the problem as a multi-class regression. In that case,  training a DNN model involves comparing its output to a source speaker. DNN models always output a dimension for each target class and when multiple sources of the same type exist,  the system needs to select arbitrarily which output dimension to map to each output and this raises a permutation problem (permutation ambiguity) \cite{Hershey2016}. Taking a case of a two speaker separation, if the model estimates $\hat{a_1}$ and $\hat{a_2}$ as the magnitude of the  reference speech magnitudes $a_1$ and $a_2$ respectively, it is unclear the order in which  the model will output the estimates i.e. the order of output can either be $\{\hat{a}_1 ,\hat{a}_2\} $ or $ \{\hat{a}_2 ,\hat{a}_1\}$.  A naive approach shown in figure 2 \cite{K2017} is to present the reference  speech magnitudes in a fixed order and hope that it is the  same  order in which the system will output its estimation.
  \begin{figure}[H]
	\centering
\includegraphics[scale=0.5,angle=0]{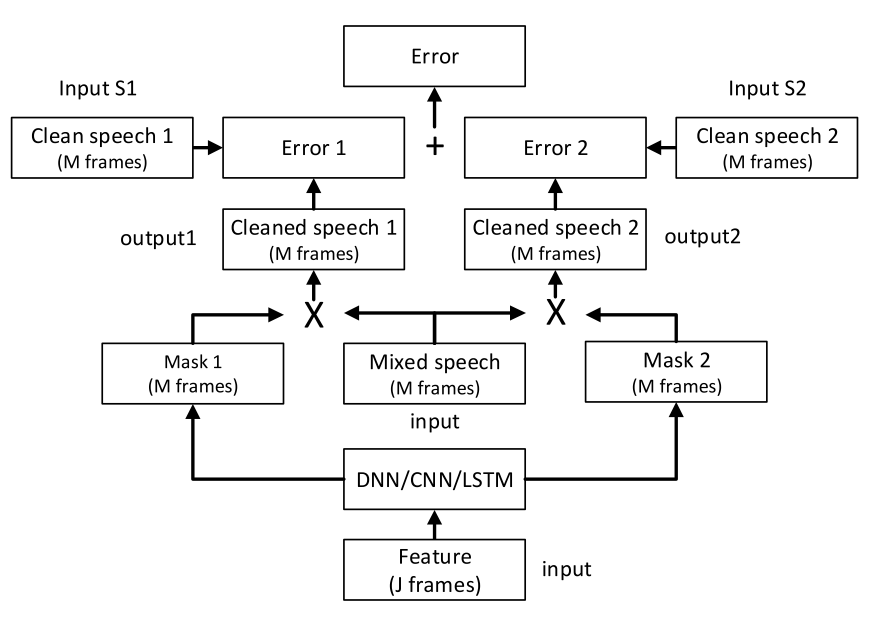}
		\caption{Naive approach of solving label matching problem for a two-talker speech separation model}
 
	\end{figure}
 In case of a mismatch, the loss computation will be based on the wrong comparison resulting in low quality of separated speeches.
Therefore, systems that perform speaker separation have an extra burden of designing  mechanisms that are geared towards handling  the permutation problem. There are several strategies that are being implemented by speech separation tools to tackle permutation problem. In \cite{Weng2015},  a number of DNN techniques are implemented that estimates two clean speeches contained in a two-talker mixed speech. They employ supervised training to train DNN models to discriminate the two speeches based on average energy, pitch and  instantaneous energy of a frame. Work in \cite{Yu2017} and  \cite{K2017} introduce permutation invariant training (PIT)  technique of computing permutation loss such that  permutations of reference labels are presented as a set to be compared with the output of the system. The permutation with the lowest loss is adopted as the correct order. For a 
a two-speaker separation system introduced earlier, the reference sources permutation will be $\{a_1,a_2\}$ and $\{a_2,a_1\}$ such that the possible permutation losses are computed as:
\begin{equation*}
    loss_1=D([a_1,a_2],[\hat{a}_1,\hat{a}_2])=D[a_1,\hat{a}_1]+D[a_2,\hat{a}_2]
\end{equation*}
\begin{equation*}
   loss_2=D([a_1,a_2],[\hat{a}_2,\hat{a}_1])=D[a_1,\hat{a}_2]+D[a_2,\hat{a}_1]
\end{equation*}
The one that returns the lowest loss between the two  is selected as the permutation loss to be minimized (see figure 3). For an $S$ speaker separation system a total of $S!$ permutations are generated.
\begin{figure}[H]
	\centering
\includegraphics[scale=0.5,angle=0]{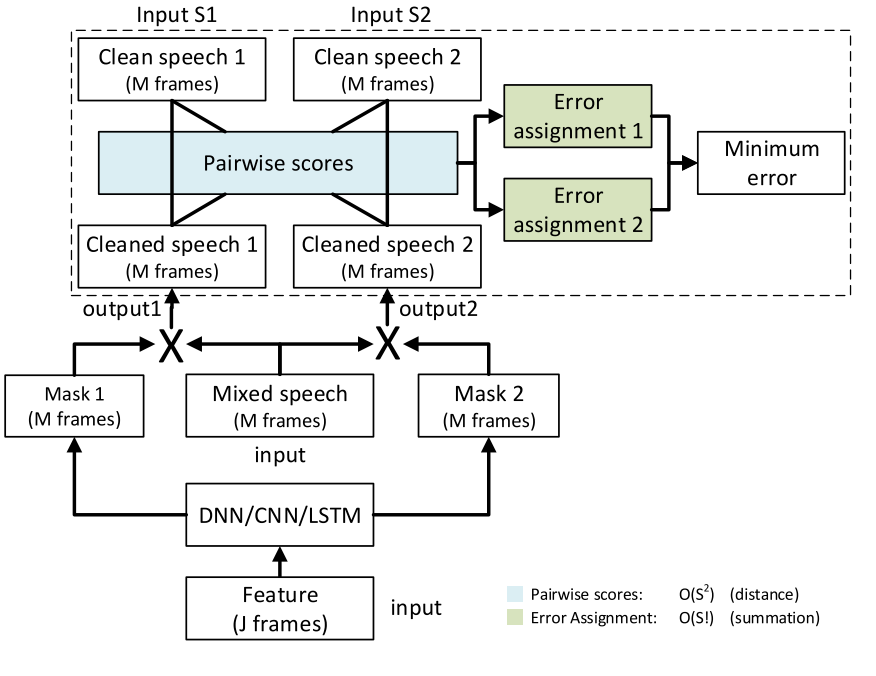}
		\caption{Permutation invariant training implementation of label matching for a two-talker speech separation model}
	\end{figure}
For a system that performs $S$ speaker separation  and $S$ is high (e.g. 10), implementation of PIT which has a computation complexity of $O(S!)$ is computationally expensive \cite{Tachibana2021} \cite{Dovrat2021}. Due to this, \cite{Dovrat2021} casts the permutation problem as a linear sum problem where Hungarian algorithm is exploited to find the permutation which minimizes the loss at computation complexity of $O(S^3)$. Work in \cite{Tachibana2021} proposes SinkPIT loss which is based on Sinkhorn's matrix balancing algorithm. They utilize the loss to reduce the complexity of PIT loss from $O(C!)$ to $O(kC^2)$. Work in \cite{Zeghidour2021} employs minimum loss permutation computation  at  each time step $t$. The best permutation (argmin) at each time-step is exploited  to re-order the embedding vectors to be consistent with the training labels.  To evade the permutation problem, they train two separate DNN models for each of the two speakers to be identified. Another prominent technique of handling permutation problem is to employ a DNN clustering technique \cite{Hershey} \cite{Byun2021} \cite{Isik2016} \cite{Qin2020} \cite{Lee2022} to identify the multiple speakers present in a mixed speech signal. The DNN $f_\theta$ accepts as its input the whole spectrogram $X$ and generates a $D$ dimension embedding vector V  i.e., $V=f_\theta(X)\in R^{N\times D} $. Here, the embedding $V$  learns the features of the spectrogram $X$ and is considered a permutation- and cardinality-independent encoding of the network’s estimate of the signal partition.  For the network $f_\theta$ to be learn how to generate an embedding vector $V$ given the input $X$, it is trained to minimize the cost function.
\begin{equation}
C_Y(V)=||VV^T-YY^T||_F^2=\sum_{ij}(<v_i,v_j>-<y_i,y_j>	)^2
\end{equation}
Here, $Y=\{y_{i,c}\}$ represents the target partition that maps the spectrogram $S_i$  to each of the $C$ clusters  such that $y_{i,c=1}$ if element $i$ is in cluster $c$ . $YY^T$  is taken here as a binary affinity matrix that represents the cluster assignment in a partition-independent way.  The goal in equation 2 is to minimise the distance between the network estimated affinity matrix $VV^T$ and the true affinity matrix $YY^T$. The minimization is done over the training examples. $||A||_F^2$ is the squared Frobenius norm. Once $V$ has been established, its rows are clustered into partitions that will represent the binary masks.  To cluster the rows $v_i$ of $V$, K-means clustering algorithm is used. The resulting clusters of $V$ are then used as binary masks to separate the sources by applying the masks on mixed spectrogram $X$. The separated sources are then reconstructed separated  by using inverse STFT. Even though  PIT is popular in speech separation models, it is unable to handle the output dimension mismatch problem where there is a  mismatch on the number of speakers between training and inference \cite{Jiang2020}. For example, training a speech separation model on $n$ speaker mixtures but testing it on $t\neq n$ speaker mixtures. The  PIT-based methods cannot directly deal with this problem due to their fixed output dimension. Most  speech separation models such as \cite{Nachmani2020} \cite{K2017} \cite{Liu2019} \cite{Luo20201} deal with the problem by setting a maximum number of sources $C$ that the model should output from any given mixture. If an inference mixture has $K$ sources, where $C> K$, $C-K$ outputs are invalid, and the model needs to have techniques to handle the invalid sources. In case of invalid sources, some models such as \cite{Liu2019}, \cite{Nachmani2020}, \cite{K2017}  design the model to output silences for invalid  sources while \cite{Luo20201} outputs the mixture itself which are then discarded by comparing  the energy level of the outputs relative to the mixture. The challenge with models that output silences for invalid sources is that they rely on a pre-defined energy threshold, which may be problematic if  the mixture also has a very low energy \cite{Luo20201}. Some models handle the output dimension mismatch problem by generating  a single speech in each iteration and subtracting it from the mixture until no speech is left \cite{Shi2018}, \cite{Kinoshita2018}, \cite{Takahashi2019}\cite{Neumann2019}, \cite{VonNeumann2020}. The iterative technique despite being trained with a mixture with low number of sources can generalize to mixtures with a higher number of sources \cite{Takahashi2019}. It however faces criticism that setting iteration termination criteria is difficult and  the separation performance decreases in later iterations  due  degradations introduced in prior iterations \cite{Takahashi2019}. Other speech  separation models include  \cite{Luo20182} \cite{Yul2017}\cite{Chang2020a} \cite{Liu2019} \cite{Chang2020} \cite{Weng2015} \cite{Isik2016} \cite{Wang2018}.
\subsection{Speaker extraction}
Some speech enhancement DNN models have been  developed where in a mixed speech such as an equation 1, they design methods to extract a single target speech. These models focus only on a single target speech $x_t$ and treat all other speeches as interfering signals, therefore they modify equation 1 as shown in 3.
\begin{equation}
    y(n)=\sum_{c=1}^{C}x_c(n)=x_t(n)+\sum_{c\not= t}^{C}x_c(n)
\end{equation}
where $x_t(n)$ is the target speech. By focusing on only a single target speech, the permutation ambiguity problem is avoided. They formulate the speech extraction task into  a binary classification problem, where the positive class is the target speech, and the negative class is formed by the combination of all other speakers.  A popular technique of speaker extraction  is  to give as input to the  DNN models  additional speaker dependent information that can be used to isolate a target speaker \cite{vesely2016sequence}. Speaker dependent information can be injected into the DNN models by either concatenating speaker dependent auxiliary clues with the input features or adapting part of the DNN model parameters for each speaker
 \cite{wang2018deep}. This addition information  about a speaker injects a bias that is necessary  to differentiate the target speaker from the rest in the mixture \cite{xiao2019single}. Several auxiliary clues have been exploited by DNN models which include pre-recorded enrolment utterances of the target speaker \cite{wang2018deep} \cite{xiao2019single} \cite{wang2018voicefilter} \cite{ji2020speaker} \cite{zhang2021towards} \cite{delcroix2018single}, electroglottographs (EGGs) of the target speaker \cite{chen2023electroglottograph} and i-vectors extracted at speaker level \cite{miao2015speaker} \cite{senior2014improving}. Tool in \cite{ochiai2014speaker} adapt  parameters for each speaker by  allocating  a speaker dependent module to a  selected  intermediate layer of DNN. Speech extraction tool in  \cite{chen2017deep} does not use auxiliary clues of the target speaker but design attractor points that are compared with the mixed speech embeddings to generate the mask used to extract the target speech.
\subsection{Dereverberation}
This is a speech enhancement technique that seeks to eliminate the effect of reverberation contained in speech. When speech is captured in an enclosed space by  a microphone that is at  distance $d$ from the talker, the observed signal consists of a superposition of many delayed and attenuated copies of the speech resulting from reflections of the enclosed space walls and existing objects within the space (see figure 4) \cite{Allen1973}. The signal received by the microphone consists of direct sound, reflections that arrive shortly after direct sound ( within approximately 50ms) i.e., early reverberation and reflections that arrive after early reverberation i.e., late reverberation \cite{Williamson2017}. Normally, early reverberation does not affect speech intelligibility much \cite{Arweiler2011} and much of  perceptual degradation of speech is attributed to late reverberation. Speech  degradation  due to reverberation can be attributed to two types of masking \cite{Nabelek1989}, overlap masking- where the energy of a preceding phoneme overlaps with the one following  or self-masking-where  internal temporal  which refers to the time and frequency alterations of an individual phoneme. 
Reverberation therefore can be viewed as  the convolution of the direct sound and the room impulse response (RIR). A reverberant speech can be formally represented according to equation 4:
\begin{figure}[H]
	\centering
\includegraphics[scale=0.5,angle=0]{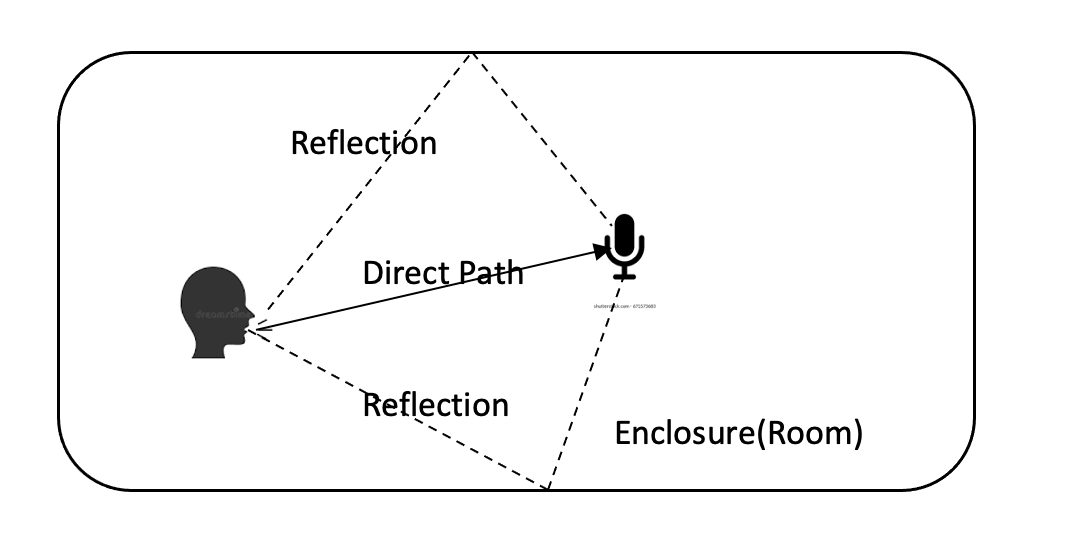}
		\caption{How Reverberation happens.}
	\end{figure}
\begin{equation}
y(t)=h(t)*s(t)
\end{equation}
Here, $*$ represents convolution, $s(t)$ is the clean anechoic speech. $h(t)$  represents room impulse response i.e., direct speech $h_d(t)$, early reverberation $h_e(t)$and late reverberation $h_l(t)$. Hence $h(t)$ can be represented as
\begin{equation}
h(t)=h_d(t)+h_e(t)+h_l(t)
\end{equation}
Using the distributive property of convolution \cite{DTSP1} equation 4 becomes:
\begin{equation}
y(t)=h_d(t)*s(t)+h_e(t)*s(t)+h_l(t)*s(t)
\end{equation}
The goal of dereverberation is therefore to establish $s(t)$ from $y(t)$.  Hence it can be viewed as a deconvolution between the speech signal and RIR \cite{Zhou2022}.
Dereverberation is considered a more challenging task than denoising for a number of reasons. First, it is difficult to pinpoint direct speech from its copies especially when the reverberation is strong. Secondly,  the key underlying assumption of sparsity and orthogonality of speech representations in the feature domain that is commonly used in monaural mask-based speech separation does not hold for speech under reverberation \cite{Cord-Landwehr2021}. Due to these unique features of reverberation,  most tools designed for denoising, or speaker separation are ill  poised to perform dereverberation \cite{Cord-Landwehr2021}. The DNN tools for speaker separation and denoising mostly make assumption that they are working on reverberation free speech hence do not make special consideration for eliminating reverberation (with exception of a few such as \cite{Su2020} \cite{Choi2020}). For instance, in \cite{Cord-Landwehr2021} they demonstrate that SepFormer \cite{Subakan2021} performance can significantly improve by making adjustments to include techniques that handle reverberation.  Several deep learning models have been designed with a goal to estimate clean speech from a reverberant one. Similar to speech  denoising and speech separation, DNN models performing dereverberation exploit these models to fit a nonlinear function to map  features of a reverberant speech to features of  clean anechoic speech  either directly \cite{Wang2019} \cite{Han2015} \cite{Jiang2014} \cite{Gamper2018} \cite{Zhao2020} \cite{Ueda2016} or by use of  mask \cite{Conference2014} \cite{Williamson2017} \cite{Williamson2017b} \cite{Jin2009} \cite{Jiang2014} \cite{Williamson2017} \cite{Jin2009}. Therefore, one way of categorising the existing dereverberation DNN tools is based on the type of target (spectrogram or ratio mask) they employ. Another way in which dereverberation tools can be categorised is based on whether a tool performs general dereverberation ( i.e suppress $h(t)$ see equation 4) or focus only on eliminating late reverberation ($h_l* s(t)$ see equation 6). Tools such as \cite{Leon2021} \cite{Zhou2022} \cite{Defossez2020} \cite{Isik2020} \cite{Li2021} \cite{Valin2022}  explore  elimination of  late reverberation. This is because early reverberation does not affect speech intelligibility much. Finally, the DNN dereverberation tools can be categorised based on  the type of training technique used (supervised or unsupervised).  Tools  such as \cite{Han2015} and \cite{Conference2014} perform speech dereverberation by implementing supervised training where the DNN model is trained to directly  estimate features clean speech when given features  from a reverberant speech. 
\begin{equation}
D(k,f)=M(k,f)\times Y(k,f)
\end{equation}
Here $D(k, f)$, $M(k, f)$and $Y (k,f)$ are the STFTs of the clean speech, the ideal ratio mask, and the reverberant speech at time frame k and frequency channel f respectively. Work in \cite{fu2022metricgan} exploits conditional GAN to perform unsupervised dereverberation of a reverberant speech.

 \textbf{Dereverberation in DTF magnitude domain}: When dereverberation is to be performed in DFT magnitude domain (see section 3.1), a DFT has to be applied to equation 4 such that,
\begin{equation}
DFT(y(t))=Y(t,f)=H(t,f)\times S(t,f)
\end{equation}
the assumption in equation  8 is that the convolution of the clean signal $s(t)$  with RIR $h(t)$ corresponds to the multiplications of their Fourier transform  in the T-F domain. However, this is only true  if the extent of $H(t,f)$ is smaller than the analysis window \cite{Cord-Landwehr2021}. Therefore, when performing dereverberation in the TF domain the selection of the window is crucial on the performance of the DNN model \cite{Cord-Landwehr2021}.\\
\textbf{Target selection in dereverberation}:In dereverberation training, most tools use direct speech as the target. This therefore means that the estimated speech will have to be compared with the direct path speech via a selected loss function. This has the potential of resulting in large prediction errors which can cause speech distortion \cite{Zhou2022}. Due to this, recent work \cite{Valin2022} proposes the use of a target that has early reverberation. By doing this, they suggest it will improve the quality of enhanced speech. In fact, experiments in \cite{Valin2022} demonstrate that allowing early reverberation in the target speech improves the quality of enhanced speech. 
\subsection{Speech denoising}
This is a speech enhancement technique of separating background noise from the target speech. Formally, the noisy speech is represented as:
\begin{equation}    
y_t=s_t+n_t
\end{equation}
where $y_t$ is the noisy speech, $s_t$ is the target speech and $n_t$ is the noise. Speech denoising seeks to isolate a single target speech from noise. Hence data driven DNN models are optimized to predict  $s_t$ from $y_t$. Since speech denoising has only a single target speech it does not suffer from global permutation ambiguity  problem. Some DNN tools that perform speech denoising include \cite{Leglaive2019} \cite{Leglaive2018} \cite{Bando2018} \cite{Leglaive2020} \cite{Kol22017} \cite{lu2022conditional} \cite{lu2021study} \cite{Lu2013} \cite{Fu2016} \cite{Gao2016}.
\section{Speech separation and enhancement features}
Speech enhancement and separation tools' input features  can be categorised into two:
\begin{enumerate}
    \item Fourier spectrum features.
    \item Time domain features.
\end{enumerate}
\subsection{Fourier spectrum features}
Speech enhancement and separation tools that use these features do not work directly on the raw signal (i.e., signal in the time domain) rather they incorporate the discrete Fourier transform (DFT) in their signal processing pipeline mostly as the first step to transform a time domain signal into frequency domain. These models recognise that speech signals are highly non-stationary, and their features vary in both time and frequency. Therefore, extracting their time-frequency features using DFT will better capture the representation of speech signal \cite{Portnoff1980}. To demonstrate the DFT process we exploit a noisy speech signal shown in equation 10. The same process can be applied in speech separation.
A noisy raw waveform signal of  speech,  $y(t)$, can be represented as in equation 10.
\begin{equation}
    y(t)=x(t)+n(t)
\end{equation}
where $x(t)$ and $n(t)$ represent discrete clean speech and noise respectively. Since speech is assumed to be statistically static for a short period of time, it is analysed frame-wise using DFT as shown in equation 11 \cite{Portnoff1980} \cite{Allen1982} \cite{Allen1977}.
\begin{equation}
    Y[t,k]=\sum_{m=\infty}^{-\infty} y(m)w(t-m)\exp^{-j2\pi km/L}
\end{equation}
Here, $k$ represents the index ( frequency bin)  of the discrete frequency, $L$ is the length of the frequency analysis and $w(n)$ is the analysis window. In speech analysis,   the Hamming window is mostly used as $w(n)$ \cite{Paliwal2011}. Once the DFT has been applied to the signal $y(t)$, it is transformed into time-frequency domain represented  as: 
\begin{equation}
    Y[t,k]=X[t,k]+N[t,k]
\end{equation}
 $Y[t,k],X[t,k]$ and $N[t,k]$ are the DFT representations of the noisy speech, clean speech and noise respectively. 
Each term in equation 12 can be expressed in terms of DFT magnitude and phase spectrum. For example, the polar form (including magnitude and phase) of the noisy signal $Y[t,k]$ is:
\begin{equation}
    Y[t,k]=|Y[t,k]|\exp^{j\angle {Y[t,k]}}
\end{equation}
$|Y[t,k]|$ and $\angle{Y[t,k]}$ are the magnitude and phase spectra of $Y[t,k]$ respectively.
Equation 13 can be written in Cartesian coordinates as shown in equation 14.
\begin{equation}
    Y[t,k]=|Y[t,k]|(\cos{\theta}+i\sin{\theta})=|Y[t,k]|\cos{\theta}+i|Y[t,k]|\cos{\theta}
\end{equation}
Both the phase and the magnitude are computed from the real and the imaginary part of $Y[t,k]$ i.e.
\begin{equation}
    |Y[t,k]|=\sqrt{\mathbb{R}(Y[t,k])^2+\Im{(Y[t,k]^2)}}
\end{equation}
\begin{equation}
    \angle{Y[t,k]}=\tan^{-1}\frac{\Im{(Y[t,k])}}{\mathbb{R}(Y[t,k])}
\end{equation}
All models that work with the Fourier spectrum features either use the DFT representations directly as the input of the model or further modify the DFT features. The features based on Fourier spectrum include:
\begin{enumerate}
   \item Log-power spectrum features.
    \item Mel-frequency spectrum features.
    \item DFT magnitude features.
    \item Complex DFT features.
    \item Complementary features.
    
    \end{enumerate}
\textbf{DFT magnitude features}: These are features  where the mixed raw waveform $y(t)$  is first converted into time-frequency (TF) representation (spectrogram) using DFT ( specifically, short-time Fourier transform (STFT) (equation 11) \cite{Natsiou2021}. The magnitude of the time-frequency representation (equation 13) acts as the input to a deep neural network (DNN) model for speech separation. The DNN model is then trained to learn how to separate the TF-bins such that those that comprise each source speech are grouped together. DNN  speech enhancements and separation  models that exploit DFT features include systems such as 
 \cite{Nossier2020} \cite{Lu2013} \cite{Grais2018} \cite{Fu2019} \cite{Jansson2017} \cite{Kim2015}.
The use of DFT magnitude as features work with high frequency resolution hence necessitating the use of larger time window which is typically more that 32ms \cite{Isik2016} \cite{K2017} for speech and more than 90ms for music separation \cite{Luo2017}. Due to this, these models must  handle increased computational complexity \cite{Baby2014}. This has motivated other speech separation models to work with lower dimensional features as compared to those of DFT magnitude.\\
\textbf{DFT complex features}: Unlike the DFT magnitude features that only use the magnitude of T-F representations, tools that use  DFT complex features include both the magnitude and the phase of the noisy (mixed) speech signal in the estimation of the enhanced or separated speech. Therefore, each T-F unit of a complex features is a complex number with a real and imaginary component (see equation 13). The magnitude and  phase of a signal is computed according to equation 15 and 16 respectively. Tools that use DFT complex features include \cite{Fu2017} \cite{Williamson2017} \cite{kothapally2022complex} \cite{kothapally2022skipconvgan}. \\
\textbf{Mel-frequency cepstral coefficients (MFCC) features}: Given the mixed speech signal such as in equation 10, to extract Mel frequency cepstral features, the following steps are executed:
\begin{enumerate}
    \item  Perform DFT of the input noisy signal  $DFT(y(t))= Y[t,k]=X[t,k]+N[t,k]$
\item Given the DFT features $Y[n,k]$  of the input signal, a filterbank with M filters i.e.  a $1\leq m\leq M$ is defined where $m$ is a triangular  filter given by:
\begin{gather}
H_m[k]=\begin{cases}
   0 & k< f[m-1]   \\
  \frac{(2(k-f[k-m])}{f[m+1]-f[m-1])(f[m]-f[m-1])}   & f[m-1]\leq k\leq f[m]\\
  \frac{2(f[m+1]-1)}{(f[m+1]-f[m-1])(f[m+1]-f[m])} & f[m]\leq k \leq f[m+1]\\
  0 & k> f[m+1]
\end{cases}
\end{gather}
The filters are used to compute the average spectrum around centre frequencies with increasing bandwidths as shown in figure 1. Here, $f[m]$ are uniformly spaced  boundary points  in the Mel-scale which is computed according to equation 18. The Mel-scale B is given by equation 19 and $B^{-1}$ which is its inverse is computed as shown in equation 20.
\begin{equation}
    f[m]=\frac{N}{F_s}B^{-1}(B(f)+m\frac{B(f_h)-B(f_l)}{M+1})
\end{equation}
 $F_s$ is the sampling frequency, $f_l$ and $f_h$ represent the lowest and the highest frequencies  of the filter bank in Hz. N is the size of DFT and  M is the number of filters. 
 \begin{equation}
     B(f)=1125\ln{(1+\frac{f}{700})}
 \end{equation} 
 \begin{equation}
     B^{-1}(b)=700(\exp{(\frac{b}{1125})-1})
 \end{equation}
 \begin{figure}[H]
	\centering
\includegraphics[scale=0.5,angle=0]{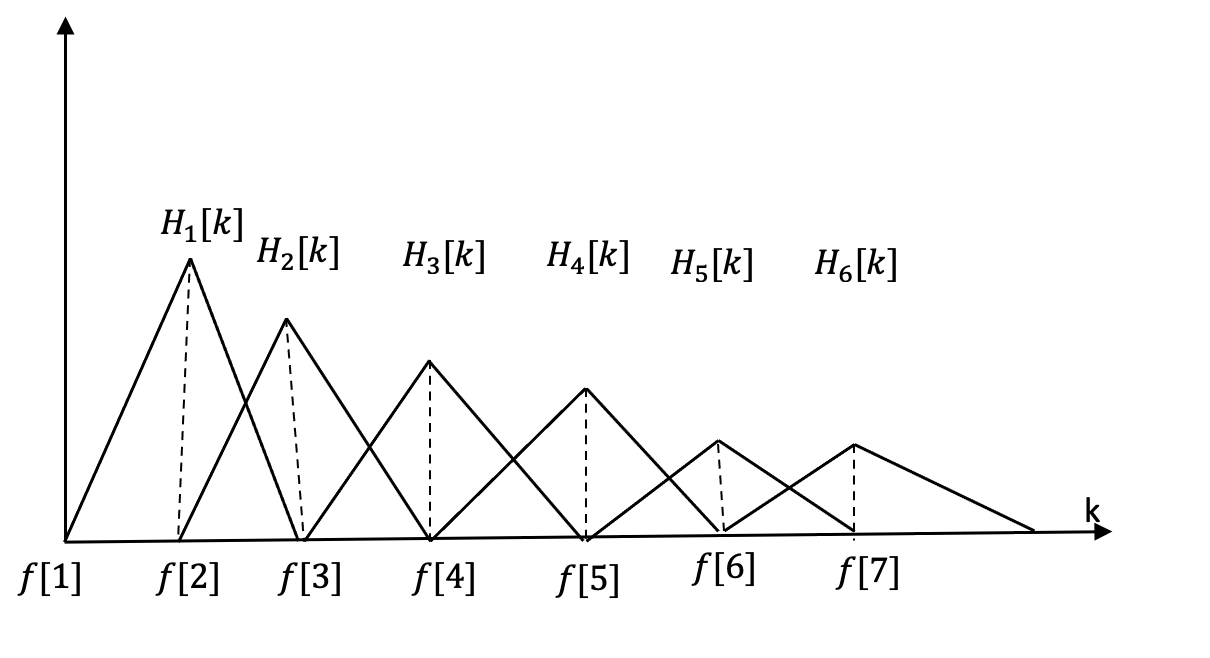}
		\caption{Triangular filters used in the computation of the Mel-cepstrum using equation 18}
	\end{figure} 
\item Scale the  magnitude spectrum $|Y[t,k]|$ of the noisy signal in both frequency and magnitude using mel-filter bank $H(k,m)$ and then take the logarithm of  the scaled frequency.
\begin{equation}
    X^{\prime}(m)=\log(\sum_{k=0}^{N-1}|Y[t,k]|^2H_m[k]
\end{equation}
for $m=0,\cdots,M $where $M$  is the number of filter banks. 
\item Compute the Mel frequency by computing  the discrete cosine transform of the $m$ filter outputs as shown in equation 22.
\begin{equation}
c[n]=\sum_{m=0}^{m-1}X^{\prime}(m)\cos{(\pi n(m+1/2)/M)}
\end{equation}
where $0\leq n< M$
\end{enumerate}
The motivation for working with MFCC is that it results in reduced resolution space as compared to DFT features. Fewer parameters are easier to learn and may generalise better to unseen speakers and noise \cite{Baby2014}. The challenge however with working on 
 a reduced resolution such as MFCC is that the DNN estimated features must be extrapolated to the DFT feature space. Due to working on a reduced resolution, MFCC degree-of-freedom will be restricted by the dimensionality of the reduced resolution feature space which is much less than that of the DFT space. The  low-rank approximation generates  a sub-optimal Wiener filter which cannot account for all the added noise content and yields reduced SDR  \cite{Baby2014}. MFCC features have been exploited in tools such as \cite{Liu2022} \cite{Ueda2016} \cite{Du2020} \cite{Lu2013}\cite{Weninger2014} \cite{Donahue2018}.\\
\textbf{Log-power spectra features}: To compute these features, a short-time Fourier analysis is applied to the raw signal computing the DFT of each overlapping  waveform (see equation 11). The log-power spectra are then computed from the output of the DFT. Consider a noisy speech signal  in the time-frequency domain i.e., where DFT has been applied to the signal (see equation 12).
From equation 14,  the  power spectrum of the noisy signal  can be represented as in equation 23.
\begin{equation}
    |Y[k]|^2=|X[k]|^2+N[k]|^2=|X[k]|^2+|N[k]|^2+2|X[k]||N[k]|\cos{\theta}
\end{equation}
Here, $\theta$ represents the angle between the two complex variables $|X[k]|$ and $|N[k]|$. Most models that exploit log-power spectra features ignore the last term ( assume the value to be zero) and  employ equation 24.
\begin{equation}
 Y^l= log (|Y[k]|^2)=log(|X[K]|^2)+log (|N[k]|^2 ) 
\end{equation}
Figure 6 \cite{Du2008} summarises the process of log-power feature extraction. 
\begin{figure}[H]
	\centering
\includegraphics[scale=0.5,angle=0]{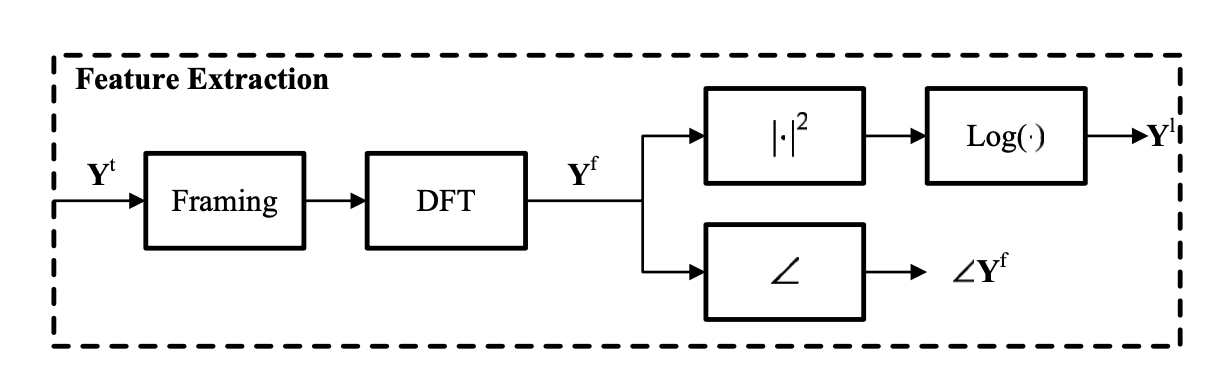}
		\caption{Demonstrating feature extraction. Here, $Y^{t}$ represents the noisy signal in time domain, 
  $Y^f$ represents the transformed signal in the frequency domain. $Y^l$ is the log power features of the input signal}
	\end{figure}
Examples of models that use Log-power spectra features include \cite{Fu2017} \cite{Du2008} \cite{Xu2015} \cite{Du2014}.\\
\textbf{Complementary features}: Since different features strongly capture different acoustic features which characterise different properties of the speech signal, some DNN models exploit a combination of the features to perform speech separation. This is based on  works such as \cite{Garau2008} and \cite{Zolnay2007} which demonstrated that complementary features significantly improve performance in speech recognition. The complementary features used in \cite{Zolnay2007} \cite{Wang20132} \cite{Williamson2016}include perceptual linear prediction, amplitude modulation spectrogram (AMS), relative spectral transform and perceptual linear product (RASTA-PLP), Gammatone frequency cepstral coefficient, MFCC, pitch-based features. The complementary features are combined by concatenation. Research in \cite{Williamson2016} reports that the use of complementary features registered better results as compared to those of DFT magnitude. The challenge  with using complementary features is how to effectively  combine  the different features, such that those complementing each other  are retained while redundant ones are eliminated \cite{Wang20132}.
\subsubsection{Supervised speech enhancement and separation training with Fourier spectrum features}
DNN models that are trained via supervised learning  using  Fourier spectrum features employ several strategies to learn how to generate estimated clean signal from a noisy (mixed) signal. These strategies can be classified into three categories based on the target of the model.
\begin{enumerate}
    \item  Spectral mapping  techniques.
    \item  Spectral masking techniques.
    \item  Generative  modelling.
\end{enumerate}
\textbf{Spectral mapping  techniques}\\
These  models fit a nonlinear function to learn a mapping from a mixed signal feature to an estimated clean signal feature (see figure 7).
\begin{figure}[H]
	\centering
\includegraphics[scale=0.34,angle=0]{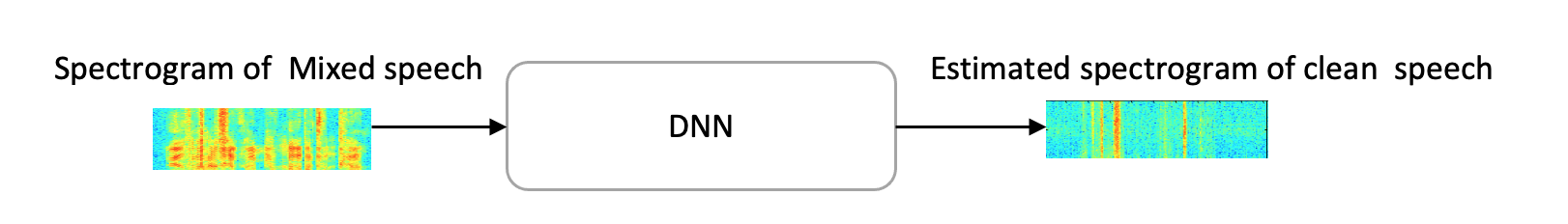}
		\caption{Supervised training of speech enhancement model  using spectrogram as  input and spectrogram as  output.}
	\end{figure}
The training dataset of these models consist of a noisy speech signal (source) and clean speech (target) features. The process of training  these models can be generalised in the following steps:
\begin{enumerate}
\item Given $N$  raw waveforms of  mixed (noisy)  speech, convert the $N$ raw waveform of noisy  speech to the desired  representation (such as spectrogram).
\item Convert the respective $N$ clean speech waveform in time domain to the same representation as that of the noisy speech.
\item Create an annotated dataset consisting of a pair of noisy speech features and that of clean speech i.e., $<noisy\_speech\_features_i, clean\_speech\_features_i>$ with $1 \leq i\leq N$
\item Train a deep learning model $g_\theta$ to learn how to estimate clean features  $clean\_speech\_features_i$ given a noisy speech feature as input $noisy\_speech\_features_i$ by minimizing an objective function.
\item Given new a noisy speech features  $x_j$  the trained model  $g_{\theta}$ should estimate a clean speech feature $y_j$.
\item Using the estimated clean speech features $y_j$, reconstruct its raw waveform by performing the inverse of the feature generation process (such as using the inverse short-time Fourier transform if the features are in time-frequency domain). 
\end{enumerate}
The above generalisation has been exploited in \cite{Lu2013} \cite{Grais2018} \cite{Kim2015} \cite{Xu2015} \cite{Lur2013} \cite{Xu2014} \cite{Fu2016} \cite{Gao2016} to achieve speech enhancement and in \cite{Jansson2017}and \cite{Weninger2014} to perform speech separation and enhancement. Figure 8 gives a summary of the steps when time-frequency(spectrogram) is exploited as the input of the speech enhancement model.

\begin{figure}[ht]
\centering
\includegraphics[scale=0.34,angle=0]{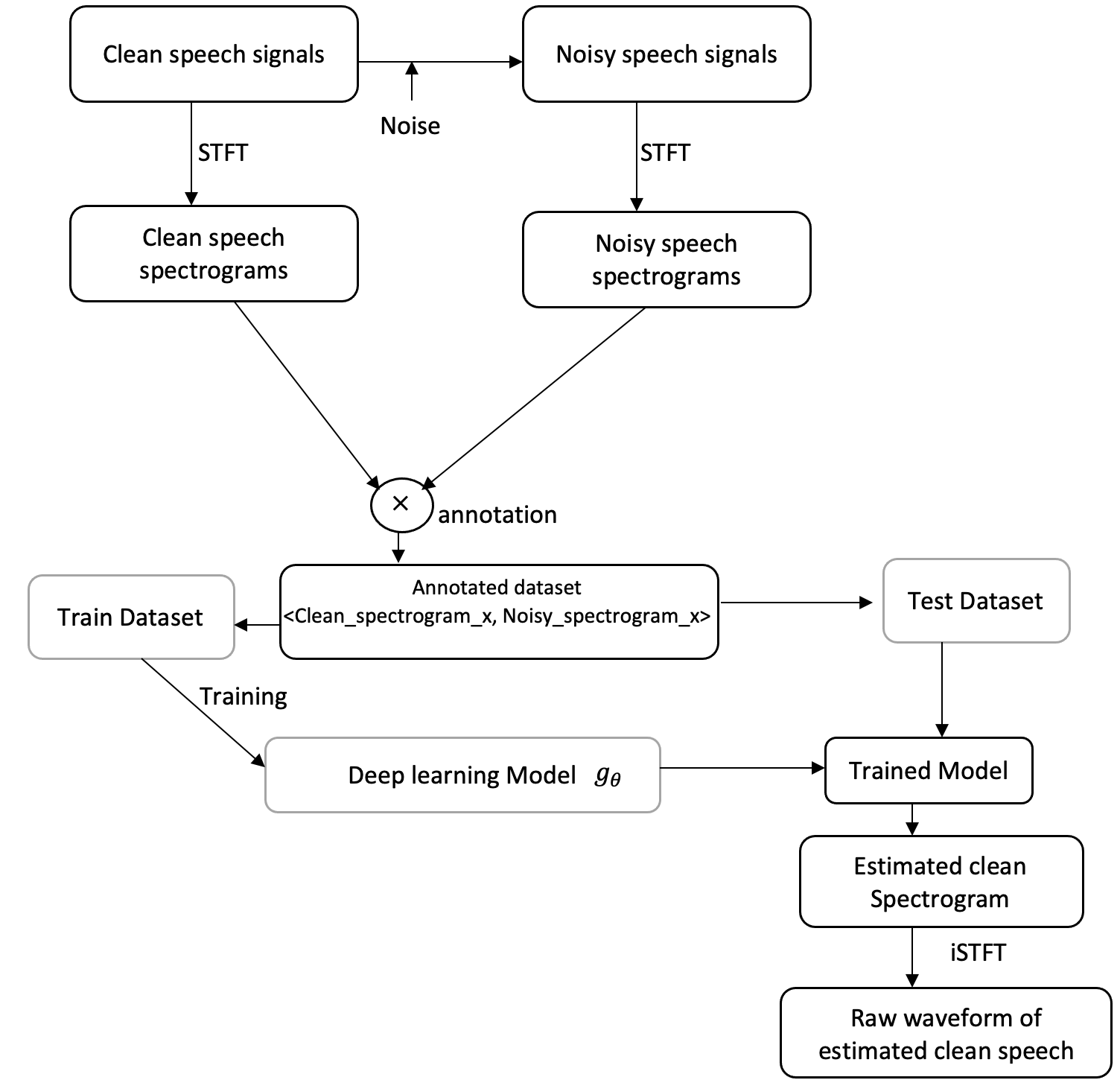}
		\caption{Showing steps involves to train speech enhancement model to fit a regression function from noisy spectrogram to an estimated clean speech spectrogram}
	\end{figure}
\textbf{Spectral masking techniques}\\
Here, the task of estimating clean speech features from  a noisy (mixed) speech input features is formulated as that of predicting real-valued or complex-valued masks \cite{Williamson2017}. The mask function is usually constrained to be in range the [0,1] even though different types of soft masks have been proposed (see \cite{K2017} \cite{Narayanan2013} \cite{Erdogan2015}). Source separation based on masks is predicated on the assumption of sparsity and orthogonality of the sources in the domain in which the masks are computed \cite{Cord-Landwehr2021}. Due to the sparsity assumption, the dominant signal at a given range (such as time-frequency bin) is taken to be the only signal at that range (i.e. all other signals are ignored except the dominant signal). In that case, the role of  DNN estimated  mask is to estimate the dominant source at a given range. To do this, the mask is applied on the input features such that it eliminates portion of the signal( where the mask has a value of 0) while allowing others (mask value of 1)\cite{Kjems2009} \cite{Wang2008}. The masks are always established by computing the signal-to-noise (SNR) within each TF bin against a threshold or a local criterion \cite{Kjems2009}. It has been demonstrated experimentally that the use of masks significantly improves speech intelligibility when an original speech is composed of noise or a masker speech signal \cite{Brungart2006} \cite{Nossier}. For deep learning models working on the time-frequency domain, a model $g_{\theta}$ is designed such that given a noisy or mixed speech spectrogram $Y[t,n]$ at time frame $t$, it estimates the mask $m_t$  at that time frame.  The established mask $m_t$ is then applied to the input spectrogram to estimate target or denoised spectrogram i.e., $\hat{S}_t=m_t \otimes Y[t,n]$ (see figure 9). Here, $\hat{S}_t$ is the spectrogram estimate of the clean speech at time frame $t$ and $\otimes$ denotes element wise multiplication. To train the model $g_{\theta}$, there are two key objective variants. The first type  minimizes an objective function $D$  such as  mean squared error (MSE)  between the model estimated mask $\hat{m}_m$ and the target mask ($tm$).
\begin{equation*}
\mathcal{L}=\sum_{u,t,f}D|tm_{u,t,f},\hat{m}_{u,t,f}|
\end{equation*}
This approach however cannot effectively handle silences where $|Y[t,n]|=0$ and $|X[t,n]|=0,$ because the target masks $tm$ will be undefined at the silence bins. Note that target masks such ideal amplitude mask (IRM) that is defined as $IRM(t,f)=\frac{|X_s(t,f)|}{\sum_{i=1}^{s}|Y_s(t,f)|}$ involves division of $|X[t,n]|$ by $|Y[t,n]|$ hence silence regions will make the target mask  undefined \cite{K2017}. This cost function also focuses on minimizing  the disparity between the masks instead of the features of estimated signal and the target clean signal \cite{K2017}. The second type of cost function seeks to minimize the features of estimated signal $\hat{S}_t=m_t \otimes Y[t,n]$  and those of target clean signal $S$  directly as shown equation 25.   
\begin{equation}
\mathcal{L}=\sum_{u,t,f}D|\hat{m}_{u,t}Y_{u,t,f},S_{u,t,f}|
\end{equation}
 The sum is over all the speech $u$ and time-frequency bin $(t,f)$. Here, $Y$ and $S$ represents noisy (mixed) and clean (target) speech respectively. So, for DNN tools using indirect estimation of clean signal features, instead of them estimating the clean features directly from the  noisy features  input, the models first estimate binary masks. The binary masks are then applied to the noisy features to separate the sources (see figure 5, here, the features are the TF spectrogram).  This technique has been applied in \cite{Wang2013} \cite{Isik2016} \cite{Weninger2014} \cite{Fu2016} \cite{Narayanan2013} \cite{Electric2015} \cite{Huang2015} \cite{Hershey} \cite{Tuzla2014} \cite{Zhang2016} \cite{Narayanan2015} \cite{Weninger2015} \cite{Conference2014} \cite{Zhang2016} \cite{Liu2019}.
 \begin{figure}[H]
	\centering
\includegraphics[scale=0.34,angle=0]{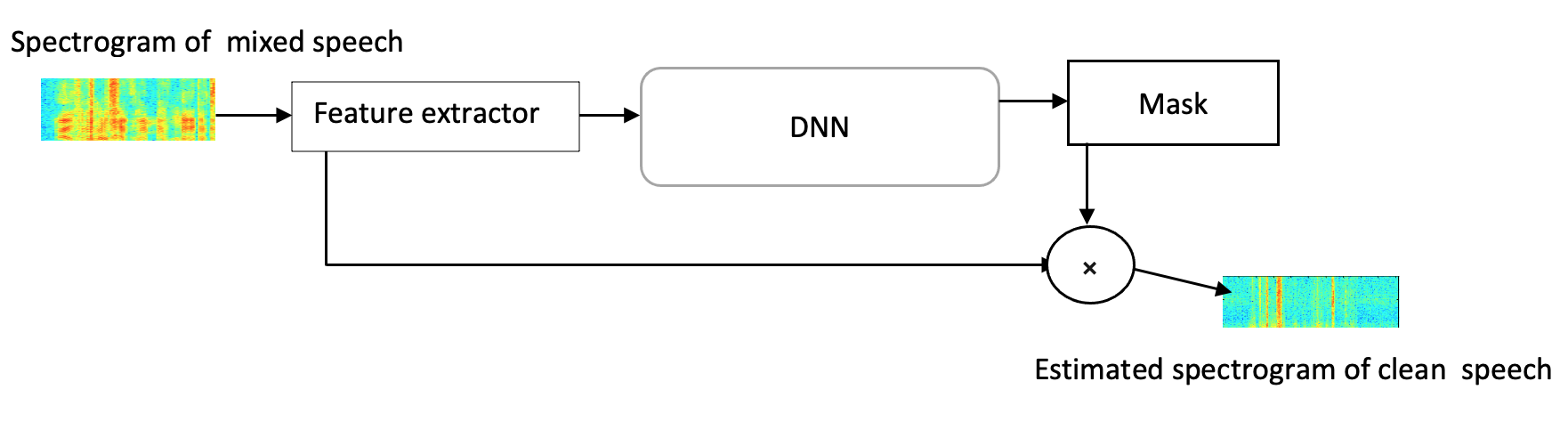}
		\caption{DNN model for mask estimation from a noisy spectrogram.}
	\end{figure}
 \textbf{Generative modelling}\\
 Given an observed sample $x$, the goal of a generative model is to model its true  distribution $p(x)$. The established model can then be used to generate new samples that are similar to the observed samples $x$. In speech separation and enhancement, these models have been exploited almost exclusively to perform speech denoising. Several generative models have been employed in a supervised manner to generate clean speech from a noisy one. Generative adversarial network (GAN) \cite{Goodfellow2016} is an unsupervised  model  that constitutes two key parts: the generator $\mathcal{G}$ and the discriminator $\mathcal{D}$, where $\mathcal{G}$  generates samples which are then  judged by the $\mathcal{D}$. The generator $\mathcal{G}$  generates the synthetic data by sampling from a simple prior $z \sim p(z) $ and the outputs a final sample $g_{\theta}(z)$ where $g_{\theta}$ is non-linear function more specifically a DNN. The  discriminator $\mathcal{D}$ on the other hand must be able to catch synthetic data as fake and real data from $p(x)$ as real. The training  objective is shown in equation 26.
\begin{equation}
 \min_{\mathcal{G}} \max_{\mathcal{D}}=
\mathbb{E}_{x\sim p(x)}[\log D(x)] -\mathbb{E}_{z\sim p(z)}[\log (1 - D(G(z)))]
\end{equation}
The objective function in equation 26 is maximised w.r.t to $D(.)$ and minimised w.r.t $G(.)$. In speech enhancement, GAN was first introduced by SEGAN( see section 3.2.3). SEGAN which works in time-domain, uses a conditioned version of the objective function of equation 26. In conditioned GAN, both the generator and the discriminator are given extra information. This allows GAN to perform classification and mapping. SEGAN uses the least-squares GAN loss  as opposed to  sigmoid cross-entropy loss used for training.  CGAN \cite{Donahue2018} just like SEGAN  uses conditioned GAN but works in T-F domain to generate a denoised speech. Since most automatic speech recognition (ASR) tools work in T-F domain, CGAN hypothesises that the generative model working in T-F domain will be more robust for ASR as compared to those working in raw waveform. Therefore,  CGAN can be seen as version of SEGAN that accepts input in T-F domain. To address the problem of mismatch between the training objective used in CGAN  and the evaluation metrics, MetricGAN \cite{Fu2019}  proposes to integrate evaluation metric in the discriminator. By doing this, instead of the generator giving  a false (0) or true (1) discrete values, it will generate continuous values based on the evaluation metric. MetricGAN can therefore be  trained to generate data according to the the selected metric score. Through this modification, MetricGAN produces more robust enhanced speech. Another common generative group of generative models is  variational auto-encoder (VAE) technique \cite{Kingma2014}. Like GAN, VAE is mainly used for denoising i.e. where the mixture is modelled as:
\begin{equation}
    x_{fn}=\sqrt{g_n} s_{fn}+b_{fn}
\end{equation}
Here, $x_{fn}$ denotes the mixture at the frequency index $f$ and the time-frame index $n$,  $g_n \in R_{+}$ is a frequency independent but frame dependent gain while $s_{fn}$ and $b_{fn}$ represent the clean speech and the noise respectively at the  frequency index $f$ and the time-frame index $n$.
We first give a brief overview of VAE before we discuss how it is adapted for speech enhancement. Mathematically, given an observable sample $s$, the goal of a generative VAE  model is to model true data distribution $p(s)$. To do this, VAE assumes that the observed sample $s$ are generated by associated latent variable $z$ and their joint distribution is $p(s,z)$. The model therefore seeks to learn how to maximize the likelihood $p(s)$ over all observed data.
\begin{equation*}
   p(s)=\int p(s,z)dz 
\end{equation*}
Integrating out all the latent variables $z$ in the above equation is intractable. However, using Evidence lower bound (ELBO) which quantifies the log-likelihood of observed data $p(s)$ can be estimated. ELBO is given in equation 28 (refer to \cite{lu2022conditional} to see  derivation of relationship between $p(s)$ and ELBO).
\begin{equation}
    \log p(s)\geq \mathbb{E}_{q_\phi(z|s)} [\log \frac{p(s,z)}{q_\phi(z\mid s)}]
\end{equation}
Here, $q_\phi(z\mid s)$ is a flexible variational distribution  with parameters $\phi$ that the model seeks to maximize. Equation 28 can be written as equation 29 using Bayes theorem.
\begin{equation}
    \log p(s)\geq \mathbb{E}_{q_\phi(z|s)} [\log \frac{p_\theta(s\mid z)p(z)}{q_\phi(z\mid s)}]
\end{equation}
Equation 29 can be expanded as:
\begin{equation}
    \log p(s)\geq \mathbb{E}_{q_\phi(z\mid s)} [\log p_\theta(s\mid z)] + \mathbb{E}_{q_\phi(z\mid s)}[\frac{\log p(z)}{q_\phi(z\mid s)}]
\end{equation}
Equation 30 can be expanded as:
\begin{equation}
    \log p(s)\geq \mathbb{E}_{q_\phi(z\mid s)} [\log p_\theta(s\mid z)] + D_{KL}(q_\phi(z\mid s)\mid \mid p(z))
\end{equation}
The second term on the right of equation 31 seeks to learn the prior $p(z)$  via $q_\phi(z\mid s)$ while the first term  reconstructs data based on the learned latent variable $z$. $q_\phi(z\mid s)$ is always modelled by a DNN and referred to as encoder and the reconstruction term is another DNN referred to as decoder. Both the encoder and decoder are trained simultaneously. The encoder is normally chosen to  model a multivariate Gaussian with diagonal covariance and the prior is often selected to be a standard multivariate Gaussian:
\begin{equation*}
    q_{\phi}(z\mid x) =\mathcal{N} (z; \mu_\theta(x),\delta^2_\theta(x)I)
\end{equation*}
\begin{equation*}
   p(z) = \mathcal{N} (z; 0, I)
\end{equation*}

To estimate clean speech based on variational-autoencoder pre-training, the tools execute several techniques that can be generalised into the following steps:
\begin{enumerate}
    \item Train a model such that it can maximise the likelihood
    $p_\theta(s\mid z)$. Here, $s$ denotes the clean speech dataset that is composed of F-dimensional samples i.e $s_t\in R^F, 1 \leq t\leq T$. The variational autoencoder assumes a D-dimensional latent variable $z_t\in R^D$. The latent variable $z_t$ and the clean speech $s_t$ have the following distribution:
    \begin{equation*}
        z_t \sim\mathcal{N}(0,I_D)
        \end{equation*}
        \begin{equation*}
            s_t \sim p(s_t|z_t)
        \end{equation*}
   
  Here, $\mathcal{N}(\mu,\delta)$ denotes a Gaussian distribution with mean $\mu$ and variance $\delta$. Basically, a decoder $p_\theta(s_t\mid z_t)$ is trained to generate clean speech $s_t$ when given  the latent variable $z_t$,  the decoder is parameterized by $\theta$.  The  decoder $p_\theta(s_t\mid z_t)$ is learned by deep learning  model during training. The encoder is trained to estimate the posterior $q_\phi(z_t|s_t)$ using a DNN. The overall objective of the variational auto-encoder training is to maximise equation 32.
\begin{equation}
  p(s)= \operatorname*{argmin}_{\theta, \phi}\sum_{i=1}^{L} \log p_\theta(s\mid z^i) + D_{KL}(q_\phi(z\mid s)\mid \mid p(z) )
\end{equation}
The posterior estimator $q_\phi(z\mid s)$ is a Gaussian distribution with parameters $\mu_d$ and $\delta_d$. These parameters are to be established by the encoder deep neural network such that $\mu_d:R^F\rightarrow R$ and $\delta_d:R^F\rightarrow R_+$.
\item Set up a noise  model  using unsupervised techniques such as NMF \cite{Hien2015}. For example, in case of NMF the noise $b_{fn}$ in equation 27 can be modelled as
   \begin{equation}
   b_{bf};w_{b,f},h_{b,n} \sim \mathcal{N}(0,(W_b,H_b)_{f,n})
   \end{equation}
where $\mathcal{N}(0,\delta)$ is a Gaussian distribution with zero mean and variance of $\delta$.
\item Set up a mixture model such that $p(x\mid z,\theta_s,\theta_u)$ is maximised. Here $x$ is the noisy speech signal, $\theta_s$ are parameters from the pre-trained model in step 1 i.e $\phi$ and $\theta$. $\theta_u={g_n,(W_b,H_b)_{f,n}} $ represents the parameters to be optimised. The parameters are $\theta_u$ are optimised by appropriate Bayesian inference technique.
    
    \item Reconstruct the clean speech $\hat{s}$ such that $p(\hat{s}|\theta_u,\theta_s,x)$ is maximised  based on the parameters $\theta_u,\theta_s$ from step 1 and 3 respectively  and the observed mixed speech $x$.
\end{enumerate}
Works that exploit different versions of  variational auto-encoder technique include \cite{Leglaive2019} \cite{Leglaive2018} \cite{Bando2018} \cite{Leglaive2020} \cite{leglaive2020recurrent}. Another generative modelling technique that has been used in speech enhancement is the variational diffusion model (VDM)\cite{sohl2015deep}. VDM is composed of two processes i.e., diffusion and reverse process.
The diffusion process perturbs data to noise and the reverse process seeks to recover data from noise. The goal of diffusion therefore is to transform a given data distribution into a simple prior distribution mostly standard Gaussian while the reverse process recovers data by learning a decoder parameterised by DNN. Formally, representing true data samples and latent variables as $x_t$ where $t=0$ represents true data and $1\leq t\leq T$ represents a sequence of latent variables, the VDM posterior is represented as:
\begin{equation}
    q(x_{1:T}\mid x_0)=\prod_{t=1}^{T} q(x_t\mid x_{t-1)}
\end{equation}
The VDM encoder $q(x_t\mid x_{t-1})$ unlike that of  VAE,  is not learned rather it is a predefined linear Gaussian model. The Gaussian encoder is parameterized with mean $u_t(x_t)=\sqrt{\alpha_t}x_{t-1}$ and variance $\varepsilon_t=(1-\alpha_t)I$. Therefore, the encoder $q(x_t\mid x_{t-1})$  can mathematically be represented as
\begin{equation}
   q(x_t\mid x_{t-1})=\mathcal{N}(x_t; \sqrt{\alpha_t}x_{t-1},(1-\alpha_t)I).
\end{equation}
$\alpha_t$ evolves over time such that the final distribution of the latent $p(x_T)$ is a standard Gaussian. The reverse process seeks to train a decoder that starts from the standard Gaussian distribution $p(x_T)$. Formally the reverse process can be represented as:
\begin{equation}
    p(x_{0:T})=p(x_T)\prod_{i=1}^{T}p_\theta(x_{t-1}\mid x_t)
\end{equation}
Here $p(x_T)=\mathcal{N}(x_T; 0,I)$. The reverse process seeks to set up a decoder $p_\theta(x_{t-1}\mid x_t)$ that optimizes the parameter $\theta$ such that:
 the conditionals $p_\theta(x_{t-1}\mid x_t)$ are established. Once the VDM is optimized, a sample from the Gaussian noise $p(x_T)$ can iteratively be
denoised through  transitions $p_\theta(x_{t-1}\mid x_t)$ for T steps to generate a simulated  $x_0$. Using reparameterization trick, $x_t$ in equation 34 can be rewritten as:
\begin{equation}
    x_t=\sqrt{\alpha_t}x_{t-1}+\sqrt{1-\alpha_t}\epsilon 
\end{equation}
where $\epsilon \sim \mathcal{N}(\epsilon,O,I)$
Similarly, 
\begin{equation}
    x_{t-1}=\sqrt{\alpha_{t-1}}x_{t-2}+\sqrt{1-\alpha_{t-1}}\epsilon 
\end{equation}
Based on this and through iterative derivation of equation 34, it can be shown that:
\begin{equation}
    x_t=\sqrt{\bar{\alpha_t}}x_0+\sqrt{1-\Bar{\alpha_t}}\epsilon_0
\end{equation}
In the  reverse process in equation 36, the transition probability $p_\theta(x_{t-1}\mid x_t)$ can be represented by two parameters $\mu_\theta$ and $\delta_\theta$ as
$\mathcal{N}(x_{t-1}; \mu_\theta(x_t,t),\delta_\theta(x_t,t)^2I)$ with $\theta$ being the learnable parameters. It has been shown in \cite{luo2022understanding} that $\mu_\theta(x_t,t)$ can be established as:
\begin{equation}
    \mu_\theta(x_t,t)=\frac{1}{\sqrt{\alpha_t}}x_t-\frac{1-\alpha_t}{\sqrt{1-\bar{\alpha_t}}\sqrt{\alpha_t}}\epsilon_\theta(x_t,t)
\end{equation}
Based on equation 40, to estimate $\mu_\theta(x_t,t)$ the DNN $\epsilon_\theta(x_t,t)$ needs to estimate the Gaussian noise $\epsilon$ in $x_t$ which was injected during the diffusion process. Like VAE, VDM uses ELBO objective for optimization. Please see \cite{luo2022understanding} for a thorough discussion on VDM. In speech denoising, work in \cite{lu2022conditional} uses conditional diffusion process to model the encoder $q(x_t\mid x_{t-1})$. In conditional encoder, instead of $q(x_t\mid x_{t-1})$, they define it as $q(x_t\mid x_0,y)$ i.e., $q(x_t\mid x_0,y)=\mathcal{N}(x_t;(1-m_t)\sqrt{\bar{\alpha}}x_0+m_t\sqrt{\bar{\alpha}}y,\delta_tI)$. Here $x_0$,$y$ represents the clean speech and noisy speech respectively. The encoder is modeled as a linear interpolation between clean speech $x_0$ and the noise speech $y$ with interpolation ratio $m_t$. The reverse process $p_\theta(x_{t-1}\mid x_t)$ is also modified to $p_\theta(x_{t-1}\mid x_t,y)=\mathcal{N}(x_{t-1};\mu_\theta(x_t,y,t),\delta I)$. Here, $\mu_\theta(x_t,y,t)$ is the mean of the conditional reverse process. similar to equation 40, $\mu_\theta(x_t,y,t)$ is estimated as
\begin{equation}
    \mu_\theta(x_t,y,t)=c_{xt}x_t+c_{yt}y-c_{\epsilon t}\epsilon_\theta(x_t,y,t)
\end{equation}
where $\epsilon_\theta(x_t,y,t)$ is a DNN model to estimate the combination of  Gaussian and non-Gaussian noise. The coefficients $c_{xt}$, $c_{yt}$ and $c_{\epsilon t}$
are established  via  the ELBO optimization. Other generative modelling techniques for speech enhancement(denoising).
\subsubsection{Highlights on Fourier spectrum features}
\begin{enumerate}
\item When performing a DFT on the input signal, an optimum window length must be selected. The choice of the window has a direct impact on the frequency resolution and the latency of the system. To achieve good performance,  most systems use 32ms. This may limit the use of the DFT based models  in environments which require short latency \cite{Luo2018}.
\item   DFT is a generic method for signal transformation that may not be optimised  for waveform  transformation in speech separation. It is therefore important to know to what extent does it place an upper bound on the performance level of speech enhancement techniques.
\item Accurate reconstruction of estimated  clean speech from the estimated features  is not easy and the erroneous reconstruction of clean speech places an  upper bound on the accuracy of the reconstructed audio.
\item Perhaps the biggest challenge when working in the frequency domain is how to handle the phase. Most DNN models only use the magnitude spectrum of the noisy signal to train the DNN then factor in the phase of the noisy signal during reconstruction. Recent works such as \cite{Paliwal2011}  have shown that this technique does not generate optimum results.
\item While working in the frequency domain, experimental research has demonstrated that spectral masking  generates better results in terms of enhanced speech quality as compared to the spectral mapping method \cite{Nossier}. 
\end{enumerate}
\subsubsection{Handling of phase in frequency domain}
The assumption made by most DNN models that use Fourier spectrum features  is that phase information is not crucial for human auditory. Therefore,  they exploit only the magnitude or power of the input speech to train the DNN models to learn  the magnitude spectrum of the clean signal and factor in the phase  during the reconstruction of the signal( see figure 10)\cite{Xu2014} \cite{Kumar2016} \cite{Du2008} \cite{Tu2014} \cite{Li2017}. The use of the phase from the noisy signal to estimate the clean signal is based on works such as \cite{Ephraim1984} that demonstrated that the optimal estimator of the clean signal is the phase of the noisy signal. Further, most speech separation models work on frames that are of size between 20-40 ms and  believe that the short-time phase contain low information \cite{Lim1979} \cite{Oppenheim1981} \cite{Vary1985} \cite{Wang1982} and therefore not crucial when estimating clean speech. However, recent research \cite{Paliwal2011} have demonstrated through experiments  that  further improvements in quality of estimated clean speech can be attained by processing both the short-time phase and magnitude spectra. Further, the factoring in of the noisy input phase during reconstruction has been noted to be a problem since the phase errors in the input interact with the amplitude of the estimated clean signal hence causing the amplitude of the estimated clean signal to differ with the amplitude of the actual clean signal being estimated \cite{Erdogan2015}, \cite{Han2015}.
\begin{figure}[ht]
	\centering
\includegraphics[scale=0.5,angle=0]{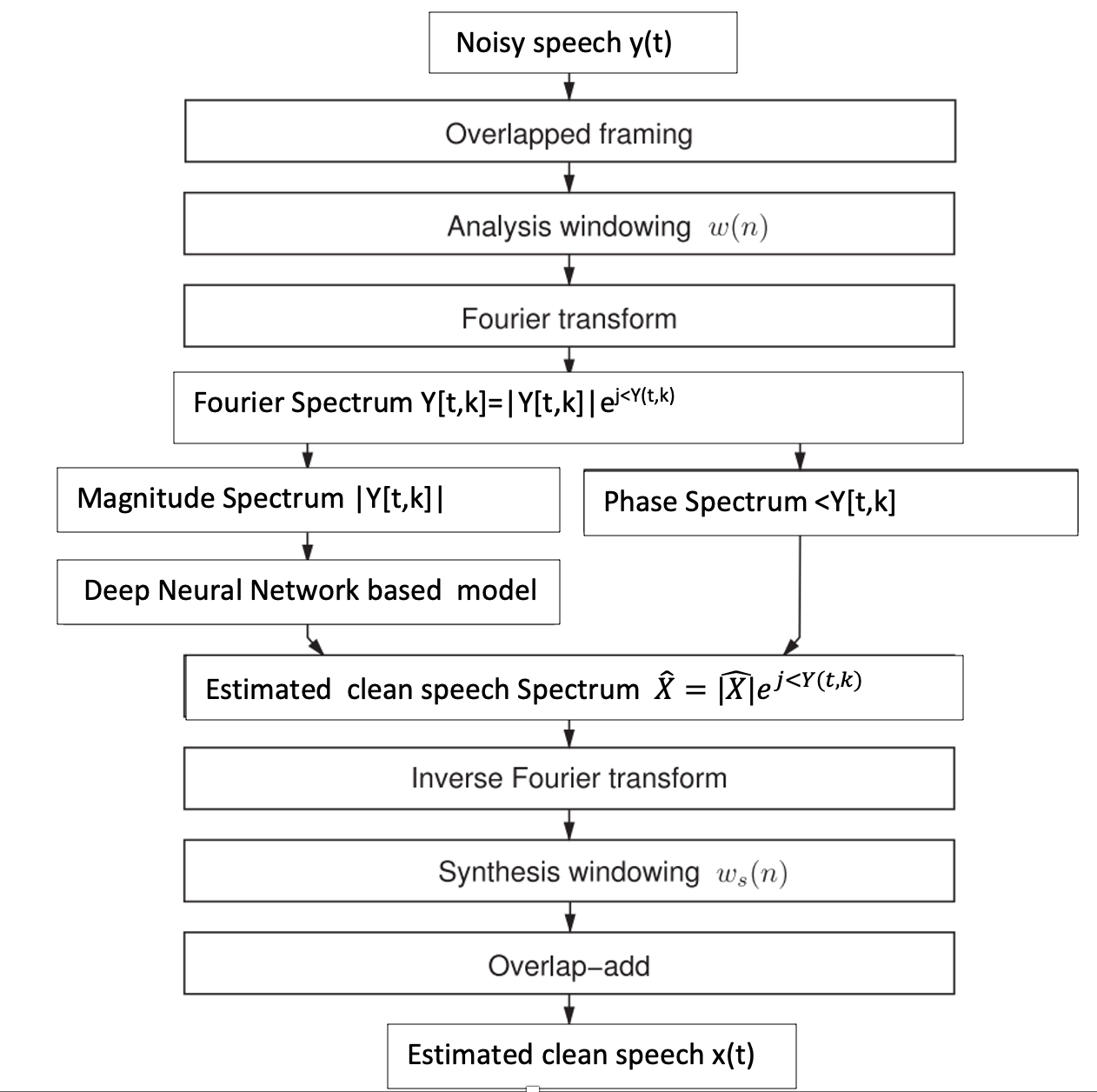}
		\caption{Showing how DNN models exploit the phase of the noisy signal during reconstruction of the estimated signal }
	\end{figure}
Based on the realisation of the importance of phase, some studies have avoided factoring in the phase of the noisy signal but rather exploit a modified phase to estimate the clean signal.  The existing techniques of modifying the phase by the DNN models working on Fourier spectrum features can be categorised into two:\\
\textbf{Phase learning}: These  models  make the phase part of the objective function i.e., they learn the phase of the estimated clean signal during training. To integrate the phase in the learning process works such as \cite{Erdogan2015} \cite{K2017} use a phase sensitive objective by replacing equation 42 with 43. It essentially  exploits  a phase sensitive spectrum approximation objective by minimising the distance between the raw waveform of the estimated speech and that of the  target clean speech. 
    \begin{equation}
\mathcal{L}=\sum_{u,t,f}D|\hat{m}_{u,t}Y_{u,t,f},S_{u,t,f}|
\end{equation}

    \begin{equation}
\mathcal{L}=\sum_{u,t,f}D|\hat{m}_{u,t}Y_{u,t,f},S_{u,t,f}(\cos{\theta}_y-\cos{\theta}_s)|
\end{equation}
Here $D$, is a selected objective function such as MSE, $\theta_y$ and $\theta_s$ represent the phase of the noisy and clean (target) speech respectively.  The sum is over all the speech $u$ and time-frequency bin $(t,f)$. Experiments conducted based on  the objective function in equation 43  show superior results in terms of signal-to-distortion ratio (SDR)\cite{Erdogan2015}. Work in \cite{Williamson2016} trains a DNN model to generate masks that are composed of  both the real and imaginary part (see equation 44). The complex mask will then be applied to  a complex representation of the noisy signal to generate the estimated clean signal. By learning a mask that has both the real and imaginary part, they integrate the phase as part of the learning.
\begin{equation}
    L=\frac{1}{2N}\sum_{t}\sum_{f}[(O_r(t,f)-M_r(t,f))^2+(O_i(t,f)-M_i(t,f))^2]
\end{equation}
 $O_r$ is the real part of the mask estimated by the DNN model while  $O_i$ is the imaginary part. $M_r$ is the real part of the target mask while $M_i$ is the imaginary part. $N$ is the number of frames and (t,f) is a given TF bin. The complex mask implementation has been exploited in \cite{Williamson2017} \cite{Erdogan2015}\cite{Lee2017} where the targets  are formulated in the complex  coordinate system i.e.  the  magnitude and phase are composed as part of the learning process. Work in \cite{Wang2018}  proposes a model that learns the phase during training via input spectrogram inversion (MISI) algorithm \cite{Gunawan2010}. Work in \cite{Ai2021} proposes a generative adversarial network (GAN) \cite{Goodfellow2016} based technique of learning the phase during training. Other works that learn the phase during training include \cite{Wang2019} and \cite{LeRoux2019}. Techniques that include phase as part of the training face the difficulty of  processing a phase spectrogram which is randomly distributed and highly unstructured \cite{Zheng2019}. To mitigate this problem and derive a highly structured phase-aware target masks, \cite{Zheng2019}  employs instantaneous frequency (IF)\cite{Friedman1985} to extract structured patterns from phase spectrograms.\\
 \textbf{Post-processing phase update}: The models that use this technique, train the DNN models using only the magnitude spectrum. Once the model has been trained to  estimate the magnitude spectrum of the clean signal, they iteratively update the phase of the noisy signal to be as close as possible to that of the target clean signal. The algorithm being exploited by the models performing post-processing phase update is based on the Griffin-Lim algorithm proposed in \cite{Griffin1984}. For example, in \cite{Han2015}, they exploit the magnitude $X^0$  of the target clean signal to iteratively obtain an optimal phase $\phi$ from the phase of the noisy signal ( see algorithm 1). The obtained phase is then used in the reconstruction of the estimated clean signal together with the magnitude $\hat{X}$ estimated by the DNN. The technique is also used in \cite{Zhao2019}. Techniques that implement Griffin-Lim algorithm  such as in algorithm 1 perform iterative phase reconstruction of each source independently  and may not be effective for multiple source separation where the sources must sum up to the mixture \cite{Wang2018}. Work in \cite{Wang2018} proposes  to jointly reconstruct the phase of all sources in a given mixture by exploiting their estimated magnitudes and the noisy phase using  the multiple input spectrogram inversion (MISI) algorithm \cite{Gunawan2010}. They ensure that  the sum of the reconstructed time-domain signals after each iteration must sum 
 to the mixture signal. Work \cite{Li2016} and \cite{Choi2020} also uses post-processing to update the phase of the noisy signal.
\begin{algorithm}
\caption{Iteratively updating the phase of a noisy signal}\label{alg:cap}
\begin{algorithmic}
\Require  Target clean magnitude $X^0$, noisy phase $\phi^0$, iteration N.
\State $ X \gets X^0, \phi \gets \phi^0,n\gets 1 $
\While{$n \leq N$} \do\\
\State $s^n \gets iSTFT(X,\phi)$
\State $(X^n,\phi^n)\gets STFT(s^n)$
\State $X\gets X^0$
\State $\phi\gets \phi^n$
\State $n\gets n+1$
\EndWhile\\
$s\gets s^n$
\end{algorithmic}
\end{algorithm}
\subsection{Time-domain features}
Due to the challenges highlighted in section 3.1.2 of working in the time-frequency domain, different models such as \cite{Luo2018} \cite{Luo2020} \cite{Luo2019} \cite{Venkataramani} \cite{Zhang2020} \cite{Subakan2021} \cite{Tzinis2020} \cite{Wang}\cite{Kong2022} \cite{Su2020} \cite{Lam2021} \cite{Lam20212} explore  the idea of designing a deep learning model for speech separation that accepts speech signal in the time-domain. The fundamental concept for these models is 
 to replace the DFT based  input  with a data-driven representation that is jointly learned  during model training.  The models therefore accept as their input the mixed raw waveform sound and then  generates 
 either the estimated  clean sources or masks that are applied on the noisy waveform to generate clean sources. By working on the raw waveform, these models address two key limitations of DFT based models. First, the models are designed to fully learn the magnitude and phase information of the input signal  during training \cite{Luo2020}. Secondly, they avoid reconstruction challenges faced when working with DFT  features.
 The time domain methods can broadly be classified into two categories \cite{Luo2020}.
\subsubsection{ Adaptive front-end based method}
The models in this category  can  roughly be discussed as composed of three key modules i.e., the encoder, separation and decoder modules ( see figure 11).
\begin{enumerate}
\item \textbf {Encoder}: The encoder can be regarded as an adaptive front-end which  seeks  to replace  STFT with a differentiable transform that is jointly  trained (learned)  with the separation model.  It accepts as its  input a time-domain mixture signal  then learns STFT-like representation \cite{Subakan2021} \cite{Kong2022}. By working directly with the time-domain signal, these models avoid the decoupling of the magnitude  and phase of the input signal \cite{Luo2019}. Most systems employ  1-dimensional convolution as the encoder to learn the features of the input signal. The transform generated by the encoder is then passed to the separation module.  Work in \cite{Kavalerov2019} demonstrates  that learned bases from raw data produce better results for speech/non-speech separation. 
\item \textbf{Separation module}: This module is fed by the output of the encoder. It implements techniques to identify the different sources present in the input signal.
\item \textbf{Decoder}: It accepts input from the separation module and sometimes from the encoder(for  residual implementation ). It is mostly implemented as  an inverse of the encoder in order to reconstruct the separated  signals \cite{Luo2018}  \cite{Luo2019} \cite{Subakan2021}.
\end{enumerate}

 \begin{figure}[H]
	\centering
\includegraphics[scale=0.34,angle=0]{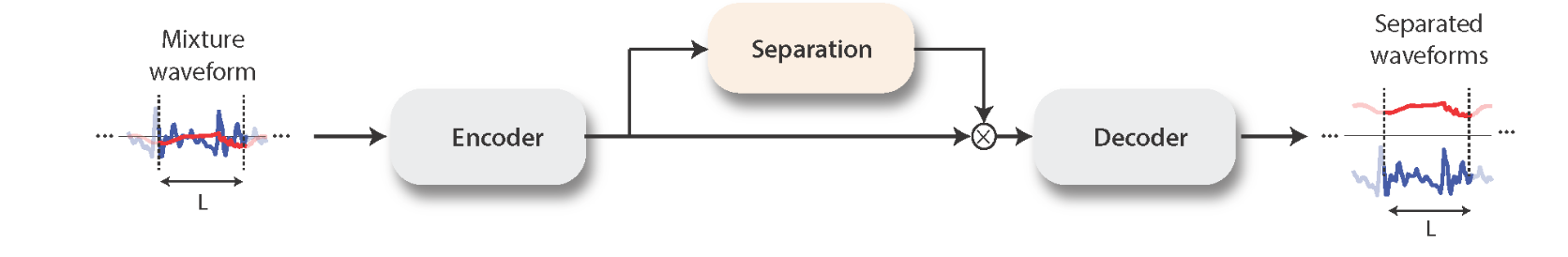}
		\caption{Showing adaptive front-end implementation.}
	\end{figure}

\subsubsection{Waveform Mapping}
The second category  of systems implement end-to-end systems where they utilise deep learning models to fit a regression function that maps an input mixed signal to its constituent estimated clean signal without an explicit front-end encoder (see figure 12). The models are trained using a pair of mixed(noisy) and clean speech. The model is fed with features of mixed signal for it to estimate clean speech. The  training involves  minimising  an objective function such as minimum mean square error(MMSE)  between the features of the clean signal and the estimated clean signal generated by the model. This approach has been implemented in  \cite{Stoller2018} \cite{Fu2018} \cite{Lluis2019}.
 \begin{figure}[H]
	\centering
\includegraphics[scale=0.34,angle=0]{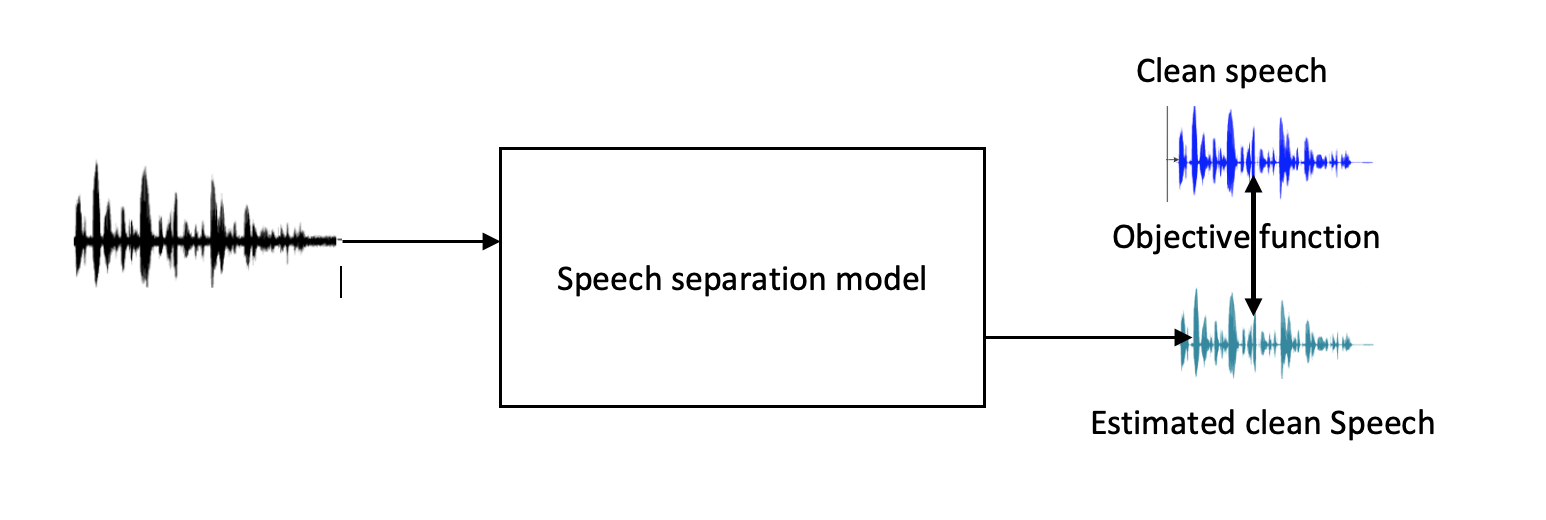}
		\caption{Direct approach training of DNN models using raw waveform.}
	\end{figure}
 \subsubsection{Generative modelling}
  SEGAN \cite{Pascual2017} is GAN based model for speech denoising that conditions   both $G$ and $D$ of equation 26 on extra information $z$ representing latent representation of the input. To solve the problem of vanishing gradient associated with optimizing objective in equation 26, they replace the cross-entropy loss by a least square function in equation 45.
\begin{equation}
 \min_{\mathcal{G}}=
\mathbb{E}_{z\sim p(z),\Bar{x}\sim p(\Bar{x})}[(D(G(z,\Bar{x}),\Bar{x})-1)^2)] +\lambda||G(z,\Bar{x})-x)||_1]
\end{equation}
Here, $\Bar{x}$ is the noisy speech, $x$ is the clean speech, $z$ is the extra input latent representation and $||.||_1$ is the $l_1$ norm distance between the clean sample x and the generated sample $|G(z,\Bar{x})$  to encourage the generator G to generate more realistic audio. Work in \cite{Pascual2019} improves  SEGAN  to handle a more generalised speech signal distortion case which involves distortions such as chunk removal, band reduction, clipping and whispered speech. Work \cite{Phan2020} improves SEGAN by implementing  multiple generators as opposed to one and demonstrates that by doing so the speech quality of the enhanced speech is better than when a single generator is used. Work in \cite{adiga2019speech} proposes a variation of SEGAN that is more tailored towards speech synthesis and  not  ASR. They replace the original loss function used in SEGAN with Wasserstein distance with gradient penalty(WGAN) \cite{gulrajani2017improved}. They also exploit gated linear unit as activation function which has been shown in \cite{Oord2016} to be more robust in generating realistic speech. Other GAN based models for speech enhancement in the time domain include \cite{xiao2021time}. Other tools that implement supervised conditional GAN include \cite{Donahue2018} \cite{Li20181} \cite{Qin2018} \cite{Fu2019}. In \cite{qian2017speech}, Bayesian network is exploited to generate estimated clean speech from a noisy one.
 \subsection{Challenges of working with time-domain features}
 \begin{enumerate}
   
 \item Time domain features lack direct frequency representation; this hinders the features from capturing speech phonetics that are present in the frequency domain. Due to this, artefacts are always introduced in the reconstructed speech in the time domain \cite{Cao2022}.
 \item The time domain waveform  has a large input space. Based on this, models working with raw waveforms are often deep and complex in order to effectively model the dependencies in the waveform. This is computationally expensive \cite{Defossez2020} \cite{Pascual2017} \cite{Subakan2021}  \cite{Wang2021}.
 \end{enumerate}.
 
\section{Which feature produces superior quality of enhanced speech?}
We performed analysis of 500 papers that exploit DNN to perform speech enhancement(i.e., multi-talker speech separation or denoising or dereverberation). We selected  papers published from 2018 to 2022. 
We were interested to answer the question, which features are more popular with these tools? The summary is presented in figure 13. Based on the analysis,  time-domain features popularity has grown rapidly from 2018 to 2022. The use of DFT features has slightly dropped, however remains popular over the five years. The popularity of  MFCC and LPS has diminished. The popularity of features that are computationally expensive such as time-domain and DFT features may be attributed to the improved computation power of computers and efficient sequence modelling techniques such as transformers and temporal convolutional networks (see section 5 for discussion). Features such as MFCC are becoming less popular due to their reduced resolution, which must be extrapolated during reconstruction hence  placing an upper bound on the quality of enhanced speech. 

\begin{figure}[H]
	\centering
\includegraphics[scale=0.5,angle=0]{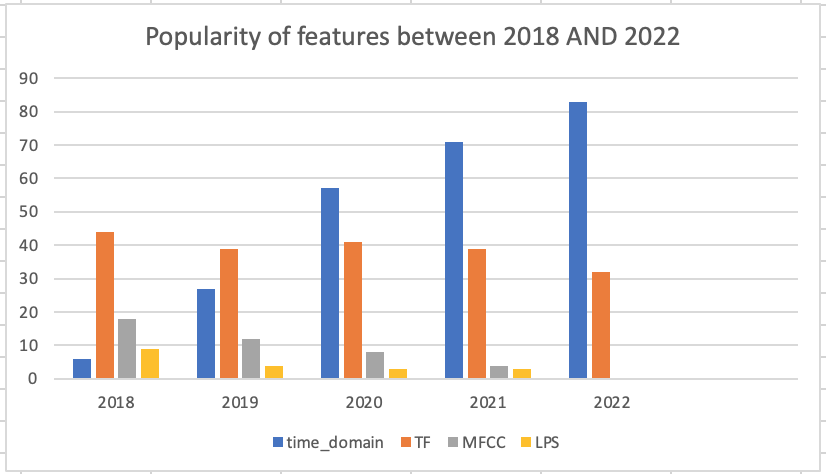}
		\caption{Feature popularity between 2018 and 2022}
	\end{figure} 
We also investigated whether  DFT or time-domain features produced the highest quality enhanced speech. Several works have conducted experiments with the goal to answer this question. Notable works include\cite{Heitkaemper2020} and \cite{Bahmaninezhad2019}. For example,  
\cite{Heitkaemper2020} investigates  Conv-TasNet's \cite{Luo2019} performance under different input types in the encoder and decoder.  Conv-TasNet uses a frame length of 4ms, stride of 2ms and overlap of 2ms. Sample results presented in \cite{Heitkaemper2020} are presented in table 1 where evaluation parameters include scale-invariant signal-to-distortion (si\_SDR), signal-to-distortion (SDR), word error rate (WER).
\begin{table}[h]
    \centering 
   \caption{Comparison of different encoder and decoder combination using Lw = 4 ms and Ls = 2ms on the test set of the WSJ0-2mix database}
     \begin{tabular}{ | c| c| c|c|c|c|}
     \hline
        \textbf{loss} & \textbf{Encoder}& \textbf{Decoder}&\textbf{si\_SDR dB} & \textbf{SDB dB}&\textbf{WER} \% \\ \hline
  $L^{SI\_SDR}$& Learned  & Learned & 14.4 & 14.7& 21.71  \\ \hline
    $ L^{SI\_SDR}$ & STFT  & Learned & 13.9 & 14.3& 21.92  \\ \hline
     $  L^{SI\_SDR}$ & Learned & iSTFT & 14.1 & 14.5& 21.87  \\ \hline
     $ L^{SI\_SDR}$ & STFT & iSTFT & 12.4 & 12.8& 24.69  \\ 
      \hline
  \end{tabular}
\end{table}
The results in table 1 show that the Conv-TasNet model gives marginally better results in terms of $si\_SDR$, $SDB dB$ and $WER$ when the input is in time domain where the signal representation is learned by the encoder and output is learned by the decoder. The results are significantly reduced in all the three parameters if STFT is used as the input and its inverse used in the decoder. For instance, Conv-TasNet model achieves a SDR of 14.7 when time-domain features are used. This  drops to 12.8 when DFT features are used. This shows that working in the time domain may be better for this setting as compared to the frequency domain. Work in \cite{Bahmaninezhad2019} also shows the same trend  where working in  time-domain provides better results as compared to frequency domain. However, for mixed speech with reverberation,  the use of a time domain signal does not improve the same results  as compared to the frequency domain and further investigation on behaviour of both time and frequency features in the presence of reverberation is needed\cite{Bahmaninezhad2019}.
\section{Long term dependencies modelling}
To effectively perform speech  separation, the speech separation tools need to model both long and short sequences within the audio signal. To do this, existing tools have employed several techniques: 
\subsection{Use of RNN}
The initial speech separation models such as \cite{Gao2016} \cite{Xu2015} \cite{non2014} relied on a feedforward DNN to estimate clean speech from a noisy one. However, feedforward DNN models are ill poised for speech data since they are unable to effectively model  long dependencies across time that are present in the speech data. Due to this, researchers progressively introduced recurrent neural networks (RNN)  which have a feedback structure such that the representations at given time step $t$ is a function of the data at time $t$, the hidden state and memory at time $t-1$. One such RNN that has been exploited in speech separation is long-short-term memory (LSTM) \cite{sema1997}. LSTM has memory  blocks that are composed of a memory cell to remember the temporary state and  several gates to control the information and gradient flow. LSTM structures can be used to model sequential prediction networks which can  exploit long-term contextual information \cite{sema1997}. Works in \cite{Weninger2014} \cite{Electric2015}\cite{Han2019} exploit LSTM to perform speech separation while \cite{Erdogan2015} uses bidirectional long short-term memory (BLSTM) networks to  make use of contextual information from both sides in the sequence. Due to their inherently sequential nature, RNN models are unable to support parallelization of computation. This limits their use when working with large datasets with long sequences due to slow training \cite{Subakan2021}. Moreover, in speech separation, a typical frame(input features) is usually 25ms which corresponds to 400 samples at a 16kHz sampling rate, for LSTM to work directly on the raw waveform, it would require unrolling the LSTM for an unrealistic large number of time steps to cover an audio of modest length \cite{Sainath2015}. Other models that use different versions of RNN include \cite{Parveen2004}. Models such as \cite{Wichern2017} use the gated recurrent unit (GRU)\cite{Cho2014} to perform speech denoising.
\subsection{Use of temporal convolution network}
Conventional convolution neural networks(CNN)have been used to design speech separation models\cite{Jansson2017} \cite{Chandna2017}. However, CNNs are limited in their ability to model long-range dependencies due to limited receptive fields \cite{Chen}. They are therefore mainly tailored to learn local features. They exploit local window which maintain translation equivariance to learn a shared position-based kernel\cite{Gulati2020}. For CNN to capture long range dependencies ( i.e., to enlarge the receptive field), there is a need to stack many layers. This increases computation cost due to the large number of parameters. These shortcomings of the CNN and RNN, have motivated the use of   dilated temporal convolution network (TCN) in speech separation  to encode long-range dependencies using hierarchical convolutional layers \cite{Rethage2018} \cite{Li2018} \cite{Lea2016} \cite{Zeghidour2021} \cite{Zhang20201}.  TCN is composed of two key distinguishing characteristics:  the convolution in the model must be causal i.e., a given activation of a certain layer $l$  at time $t$  is only influenced by activations of the previous layer $l-1$ from time steps that are less that $t$, 2) the model takes the sequence of any length and maps it into an output sequence of the same length. 
To achieve the second characteristic, TCN models are implemented using a 1-dimensional convolutional network such  that  each hidden layer is the same length as the input layer. To ensure same length, a  zero padding of length $filter size-1$  is added to keep subsequent layers the same length as previous ones \cite{Bai2018} (see figure 14).The first property is achieved through the use of causal convolutions i.e.  where an output at time $t$ is convolved only with elements from time t and earlier in the previous layer. To increase the receptive fields, models implement  dilated TCN. Dilated convolution  is  where the filter is applied to a region larger than its size \cite{He2019}. This is achieved by skipping input with certain specified steps (see figure 10). More formally, for 1D sequence such as speech signal, the input $x\in R^n$ and the kernel $f:\{0,\cdots, k-1\}\rightarrow R$, the dilated convolution operation $F$ on an element $s$ of a given sequence is defined according to equation 20 \cite{Bai2018}.
\begin{equation}
F(s)=(x\ast_d f)(s) =\sum_{i=0}^{k-1} f(i)x_{s-di}
\end{equation}
where $x$ is the $1D$ input signal, $k$ is the kernel and $d$ is the dilation factor.
 The effect of this is to expand  the receptive field without loss of resolution and drastically increase the number of parameters. Stacked dilated convolution expands the receptive field with only a few layers. The expanded receptive field allows the network to capture temporal dependence of various resolutions with the input sequences \cite{Zhang2020}. In effect, TCN  introduces the idea of time-hierarchy where the  upper layers of the network  model  longer input sequences on larger timescales  while  local information are modelled by lower  layers and are mainly maintained in the network  through  residuals and skip connections \cite{Zhang2020}. TCN also uses causal convolution where a given output at layer $l$ in time step $t$ is computed only based on time steps up to time step $t-1$ in the previous layer. The dilated TCN is exploited by \cite{Luo2019} to model sequences that exist within the input speech signal. They implement  TCN such that each layer is composed of 1-D dilated  convolution blocks. The layers have  1-D CNN blocks with increasing dilation factors. This is to uncover long range dependencies that exist in the audio input. The dilation factors increase exponentially over the layers in order to cover a large temporal context window to exploit the long-range dependencies that exist within a speech signal.
\begin{equation}
y(m,n)=\mathop{\sum_{i=1}^{M}\sum_{j=1}^{N}}x(m + r \times i, n + r \times j)w(i, j)
\end{equation}
Here, $ y(m, n)$  is the output of a given layer of  dilated convolution, $x(m, n)$ is the input and  $w(i, j)$ is the filter  with the length and the width of $M$ and $N$ respectively. The parameter $r$ is the dilation rate. Note that if $ r = 1$, the  dilated convolution  becomes the normal convolution
convolution.
\begin{figure}[H]
	\centering
\includegraphics[scale=0.5,angle=0]{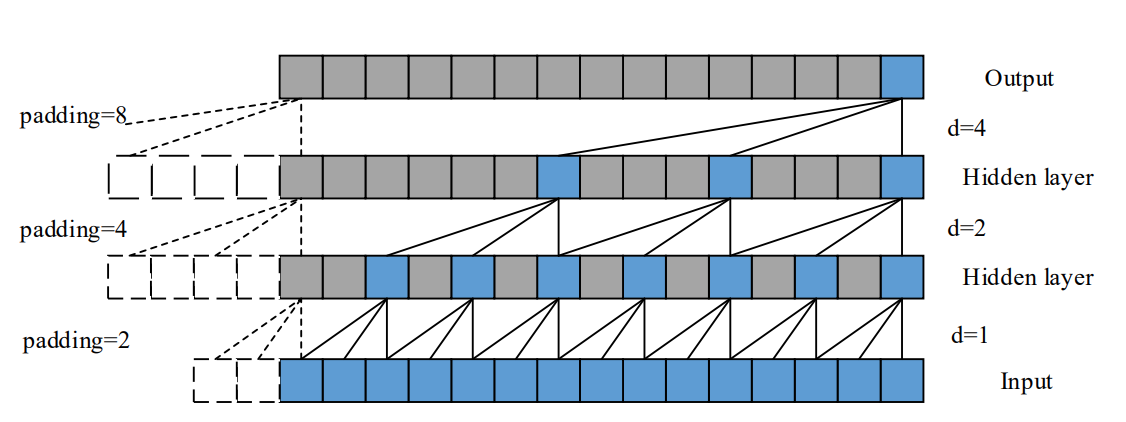}
		\caption{ Dilated TCN with four layers.}
	\end{figure}
.
\subsection{Use of transformers}
 A transformer\cite{trans2020} is an attention-based deep learning technique that has been successful in  modelling sequences and allows uncovering of  dependencies that exist within an input without regard to the distance between  any two values of the input. Transformers consist only of feed-forward layers which allows them to exploit the parallel processing capabilities of GPUs leading to fast training \cite{trans2020}.  In speech separation, \cite{Subakan2021} introduces a speech separation system that fully relies on transformers to  model the dependencies that exist in the mixed audio signal. This is used to extract a  mask for each of the  speakers in the audio mixture. The transformer is used to uncover both the short-term dependencies (within a frame) and long-term dependencies (between frames). Work in \cite{Chen} also exploits transformers in the encoder to model the dependencies that exist in the mixed audio while \cite{Zhao2020} uses transformers to perform speech  dereverberation. Despite their  ability to model long-range dependencies and ability to work  well with parallelization, the attention mechanism of transformers, has $O(N^2
)$ complexity  that brings  a major memory bottleneck \cite{Subakan2022}. For a  sequence of length $N$,  the transformer needs to compare $N^2$ elements which results in a computational bottleneck especially for long signals such as speech.  Transformers also use many parameters aggravating the memory problem further. Several versions of transformers such as  Longformer\cite{Beltagy2020}, LinFormer\cite{Wang2020}
 and Reformer\cite{Kitaev2020} have been proposed with a goal to reduce the computation complexity of the transformers. Work in \cite{Subakan20221} investigates the performance of the three versions of transformers in speech separation  and concludes  that  they  are suitable for speech separation applications since  they achieve a  highly favourable trade-off between performance and computational requirements. Work in \cite{Luo2022} proposes a technique of parameter sharing to reduce the computation complexity of the transformer while \cite{Subakan2022} reduces complexity by avoiding frame overlap. In \cite{Chen2021}, a teacher-student speech separation model based on transformer is proposed. The student model which is much smaller than the teacher model  is used to reduce computation complexity.    
 Other transformer-based speech enhancement tools include \cite{Wang2021}  \cite{DeOliveira2022}. Another key limitation of a transformer is that while it can model long-range global context, they do not extract fine-grained local features patterns well. Based on this, transformer-based speech separation tools apply attention within a frame (chunk) to capture local features and between frames(chunks) to capture global features  \cite{Subakan2021} \cite{Qiu2022}.
 \section{Model size reduction techniques}
  To achieve high performance i.e., generate speech with high intelligibility, DNN models for speech enhancements are becoming large by exploiting millions of parameters \cite{Lutati2022}. High number of parameters increase the memory requirements, computation complexity and latency.  To reduce these parameters significantly without compromising quality and make speech enhancement tools to work in resource constrained platform, several techniques are being exploited. The techniques include: \\
 \textbf{Use of dilated convolution}: To increase the receptive field of 1D CNN and subsequently increase the temporal window and model long range dependencies within a speech, speech separation such as \cite{Luo2018} and \cite{Rethage2018} implement dilated CNN. Dilated convolution initially introduced by \cite{Oord2016} involves a convolution where a  kernel is applied to an area that is larger that it. This is achieved by skipping input values by a defined step. It is like implementing  a sparse kernel ( i.e., dilating the kernel with zeros). When dilated convolution is applied in a stacked network, it enables the network to increase its receptive field with few layers hence minimizing parameters and reducing computation \cite{Lea2016}( see figure 14). This ensures that the models can capture long range dependencies while keeping the number of parameters at minimum. The dilating factors are made to increase exponentially per layer( see figure 10).\\
 \textbf{Parameter quantization}: To reduce computation, inference complexity of DNN models and to scale down the number of parameters, models such as \cite{Wu2019} \cite{Sun2017} \cite{Lin2019} \cite{Fedorov2020} \cite{Hsu2019} use parameter quantization. In quantization, the objective is to reduce the precision of model parameters and activation values to a low precision with minimal effects  on the generalization capability of the DNN model. To achieve this, a quantization operator $Q$ is defined that maps a floating value to a quantized one \cite{Gholami2022}.\\
   \textbf{Use of depthwise separable convolution}: This type of convolution, decouples the convolution process into two i.e.  depthwise convolution where a single filter is applied to each input channel and pointwise convolution which is  applied to the output of depthwise convolution to achieve a linear convolution of the depthwise layer. Depthwise separable convolution has been shown to reduce the number of parameters as compared to the convectional one \cite{Avery2014} \cite{Howard2017}. Speech enhancement tools that exploit depthwise separable convolution include \cite{Luo2019} \cite{Byun2021} \cite{Zhao2020}.\\
  \textbf{Knowledge  distillation}: Knowledge distillation involves training a large teacher model which can easily extract the structure of data then the knowledge learned by the teacher is distilled down to a smaller model called the student. Here, the student is trained under the supervision  of the teacher\cite{Hinton2015} \cite{Gou2021} \cite{Wang20221}. The student model must mimic the teacher and by doing so achieve superior or similar performance but at reduced computation cost due to reduced parameters. Knowledge distillation technique has been exploited to reduce latency in speech enhancement tool \cite{Chen2021} \cite{Aihara2019}. \\
 \textbf{Parameter pruning}: In order to reduce the number of parameters and hence speed up computation, some speech enhancement tools use parameter pruning \cite{Fedorov2020} \cite{Tan2021}.  Pruning involves converting a dense DNN model into a sparse one by significantly scaling down  the number of parameters without compromising model's output's quality. In \cite{Ye2019} they train a speech enhancement DNN model to obtain an initial parameter set $\Theta$, they then prune the parameters by dropping the weights whose absolute values are below a set pruning threshold. The sparse network is again re-trained to obtain final parameters. Work in \cite{Wu2019} estimates the sparsity $S(k)$ of a given channel $F_{jk}$. If the sparsity $S(k)>\theta$ where $\theta$ is a predefined threshold, the weights within $F_{jk}$ is set to zero and the model is retrained. After several iterations, the channel $F_{jk}$ is dropped.\\
 \textbf{Weight sharing}: This involves identifying clusters of weights that have a common value. The clusters are normally identified using K-means algorithm. So instead of storing each weight value, only the indexes of the shared values are stored. Through this memory requirements of the model is reduced \cite{Dupuis2020}. Speech enhancement tools that use weight sharing include \cite{Sun2017}, \cite{Hu2021}.

\section{Objective functions for speech enhancement and separation}
Most DNN monaural speech enhancement and separation models especially those working on features in the frequency domain exploit mean-square-error (MSE) as the training objective \cite{K2017} \cite{Luo20182} \cite{Ephraim1984} \cite{Venkataramani}. The DNN models that have the mask as the target use the training objective to minimise the MSE between the estimated mask and the ideal mask target. For models that predict estimated features (such as T-F spectrogram) of the clean source speech, MSE is used to minimise the difference between  target features and the estimated features by the model. Despite the dominance of MSE as an objective function in the speech enhancement tools, it has been criticised since it is not closely related to human auditory perception \cite{Fu2019}. Its major weakness is that it treats estimation elements independently and equally. For instance, it treats each time-frequency unit separately rather than whole spectral \cite{Xu2015}. This leads to muffled sound quality and compromises intelligibility \cite{Xu2015}. MSE also treats every estimation element with equal importance which is not the case \cite{Zhang2018}. It also does not discriminate  between the  positive or negative differences between the clean and estimated spectra. A positive difference between the clean and estimated spectra represents attenuation distortion, while a negative spectral difference represents amplification distortion. MSE treats the effects of these two distortions on speech intelligibility as equivalent which is problematic \cite{Loizou2011} \cite{P.C.Loizou2013}. Moreover,  the MSE is usually defined in the linear frequency scale while the human auditory perception is on the Mel-frequency scale. To avoid the problem of treating every estimation element with equal importance \cite{Xia2014},  \cite{Shivakumar2016} propose a weighted  MSE. Due to the shortcomings of MSE, objective functions that are closely related to the human auditory perception have been introduced to train the DNN \cite{Zhang2018} \cite{Koizumi2017} \cite{Kolbcek2018}  \cite{Yan2018} \cite{Electric2015}. Some of the human auditory perception training objectives being used by speech enhancement tools are also used as  metrics for perceptual evaluation. They include:
\begin{enumerate}
     \item Short-time objective intelligibility (STOI)\cite{Taal2011}.
    \item Scale invariant signal-to-distortion ratio(SI-SDR)\cite{Roux2019}.
    \item Perceptual metric for speech quality evaluation(PMSQE).
\end{enumerate}
\textbf{Scale invariant signal-to-distortion ratio} Work in \cite{Roux2019} proposes an intelligibility measure such that given the target signal $s$ and the model estimated signal $\hat{s}$, they re-scale  either $s$ or $\hat{s}$ such that the residual $(s-\beta \hat{s})$ after scaling $\hat{s}$ or $(\alpha s- \hat{s})$ after scaling the target $s$ is orthogonal to the target as:
$(s-\beta \hat{s}). s=0$ or $(\hat{s}-\alpha s). s=0$
based on this,$\alpha $ can be computed as:
\begin{equation*}
  (\hat{s}-\alpha s). s=0
\end{equation*}
\begin{equation*}
  \hat{s}.s -\alpha s.s =0\\
\end{equation*}
\begin{equation*}
   \alpha=\frac{\hat{s}^T s}{s.s}
\end{equation*}
based on scaling of the target $s$, the signal-to-noise ratio(SNR) equation 
\begin{equation}
    SDR=10\log_{10}\frac{||s||^2}{||s-\hat{s}||^2}
\end{equation}
is transformed to:
\begin{equation}
    SDR=10\log_{10}\frac{||\alpha s||^2}{||s-\hat{\alpha s}||^2}
\end{equation}
replacing the $\alpha$ we get the SI-SDR:
\begin{equation}
   SI-SDR=10\log_{10}\frac{||\frac{\hat{s}^T s}{||s^2||}s||^2}{||\frac{\hat{s}^T s}{||s^2||}s-\hat{s}||^2}
\end{equation}
This  objective function  has been used in \cite{Subakan2021} \cite{Luo2019} \cite{Bahmaninezhad2019} \cite{Nachmani2020}\cite{Subakan2022} \cite{Li2022}, \cite{Fan2020} \cite{Byun2021} \cite{Lee2022} \cite{Lutati2022}.

\textbf{Short-time objective intelligibility}: This objective has been used in \cite{Kolbcek2018}  \cite{Yan2018} \cite{Zhang2018} \cite{Fu2019}. STOI \cite{Taal2011}\cite{Taal2010} is a speech intelligibility measure that is achieved by executing the following steps:
\begin{enumerate}
    \item Given discrete time signals of clean speech signal $x(n)$ and enhanced speech $y(n)$, perform a DFT on both $x(n)$ and $y(n)$ i.e $X(n,k)=DFT(y(n))$ and $Y(n,k)=DFT(y(n))$. Here, $k$ refers to the index of the discrete frequency.
    \item Remove silences in both the clean signal and the enhanced signals. Silences are removed by first identifying the frame with maximum energy($max_{energy}$) in the clean signal. All frames with energy of 40 dB less than $max_{energy}$ are dropped.
    \item Reconstruct both the clean and enhanced speech signals.
    \item Perform a one-third band octave analysis on both clean and enhanced speech by grouping DFT bins i.e the  complex-valued STFT coefficients, $X(n,k)$, are  combined into J third-octave bands by computing the TF units.
    \begin{equation}
        X_j(m)=\sqrt{\sum_{k=k_1(j)}^{k_2(j)-1}|X(n,k)|^2} j=1,\cdots,J
    \end{equation}
    Here, $k_1$ and $k_2$ represent the one-third octave band edges. The same octave analysis is performed on the enhanced speech. The one-third octave of the enhanced speech is defined in a similar manner.
\item Define  a short temporal envelope  of both enhanced and clean speech as:
$Y_{j,m}=[Y_j(m-N+1), Y_j(m-N+2),\cdots,X_j(m)]^T$ and  $X_{j,m}=[X_j(m-N+1), X_j(m-N+2),\cdots,X_j(m)]^T$ respectively. STOI exploits correlation coefficients to compare the temporal envelopes of clean and enhanced speech for a short time region. Note that N=30.
\item Normalise the short temporal envelopes of the enhanced speech. Let $y_{j,m}(n)$ denote the $n^{th}$ envelope of enhanced speech. The normalised enhanced speech $y^{\prime}_{j,m}(n)$ of $Y_{j,m}(n)$ is given by $y^{\prime}_{j,m}(n)=\frac{|X_{j,m}|}{||Y_{j,m}||}Y_{j,m}(n)$.
 $||.||$ is the $l_2$ norm.
The intuition behind normalisation of the enhanced speech is to reduce global level differences between clean and enhanced speech. These global level differences should not have a strong effect on speech intelligibility.
\item Clip the normalised enhanced speech as
$\bar{y}_{j,m}(n)=\min(y^{\prime}_{j,m},(1+10^{\frac{\beta}{20}})x_{j,m}(n))$. Clipping is done to ensure the effects of severely degraded frames of the enhanced speech  on the model is upper bounded. Here, $\beta=-15dB$ is the lower signal-to-distortion(SDR) bound.
\item Compute  intermediate intelligibility measure as 
\begin{equation}
    d_{j,m}=\frac{(x_{j,m}-\mu_{x_{j,m}})^T(y_{j,m}-\mu_{y_{j,m}})}{||x_{j,m}-\mu_{x_{j,m}}||y_{j,m}-\mu_{y_{j,m}}||}
\end{equation}
Here,$\mu_{(.)}$ refers to the sample mean of the corresponding vector.
\item Compute the average intermediate intelligibility of all frames as
\begin{equation}
    d=\frac{1}{JM}\sum_{j,m}d_{j,m}
\end{equation}
where $M$ represents the total number of frames and $J$ the number of one-third octave band.
\end{enumerate}
\textbf{Short-time spectral amplitude mean square error}.
Let $X[n,k]$ with $1\leq n\leq N$ and $1\leq k\leq K$ be an $N$ point DFT of $x$ and $K$ is the number of frames.Let $A[n,k]=X[n,k]$ with $k,\cdots,\frac{N}{2}+1$ and $k=1,\cdots,K$ denote the single sided amplitude spectra of $X[n,k]$. Let $\hat{A}[n,k]$ be an estimate of $A[n,k]$, the short-time spectral amplitude mean square error(STSA-MSE) is given by
\begin{equation}
   \mathcal{L}_{STSA-MSE} =\frac{1}{(N/2+1)K}\sum_{n=1}^{N=K/2+1}\sum_{k=1}^{k=K}(\hat{A}[n,k]-A[n,k])^2
\end{equation}
Equation 36 represents the mean square error between single-sided amplitude spectra of the clean speech $x$ and the DNN estimated speech. Equation 36 is not sensitive to the phase spectrum of the two signals 
$\hat{x}$\\
\textbf{Perceptual metric for speech quality evaluation(PMSQE)}. This is an objective function that is based on the adaptation of perceptual evaluation of speech quality (PESQ) algorithm \cite{Rix2001}. Given the MSE loss in the log-power spectrum with mean and variance normalisation i.e.
\begin{equation}
    MSE_t=\frac{1}{k}\sum_{n=1}^{K}(\frac{\log |x[n,k]|^2-\mu_k}{\delta_k}-\frac{\log|\hat{X}[n,k]|^2 -\mu_k}{\delta_k})^2=\frac{1}{K}\sum_{k=1}^{k=K}\frac{1}{\delta_k^2}(\log\frac{ |X[n,k]|^2}{|\hat{X}[n,k]|^2})^2
\end{equation}
Here, $X[n,k]|^2$ and $\hat{X}[n,k]|^2$ represent the power spectra of the clean  and enhanced  speech respectively. $\mu_k$ is the mean log-power spectrum and $\delta_k$ is its standard deviation. The indices n and k represent the frame and frequency, while $K$ is the number of frequency bins. From equation 55, the MSE is entirely dependent on 
power spectra across frequency bands hence not factoring in the  perceptual factors such as  loudness difference, masking and
threshold effects\cite{Martin-Donas2018}. To factor in  the perceptual factors in the MSE, PMSQE modifies the  MSE loss by incorporating two disturbance terms (symmetrical disturbance and asymmetrical disturbance) which are based on the  PESQ algorithm both computed in a frame-by-frame basis\cite{Martin-Donas2018}.
\begin{equation}
    MSE_t=\sum_{t}{}MSE_t+ \alpha D_t^s+\beta D_t^a
\end{equation}
Here $D_t^s$ and $D_t^a$ represent symmetrical and asymmetrical disturbances respectively. The parameters  $\alpha$ and $\beta$ are weighting factors  which are determined experimentally. Work in \cite{Martin-Donas2018} describes how to arrive at the values of $ D_t^s$ and $D_t^a$.
Since  PESQ is non-differentiable, the PMSQE  objective function provides a way of estimating it. PMSQE objective function is designed to be inversely proportional to PESQ, such that a low PMSQE value corresponds to a high PESQ value and vice versa. The key question here is: Which objective function is superior?
 work in \cite{Kolbaek2020} tries to answer this question where they evaluate six objective functions. Their conclusion is that the evaluation metric should be a major factor in deciding on the objective function to use in the speech enhancement model. In case a given  model targets to improve a specific evaluation metric, then selection of an objective function related to that metric is advantageous. 
\section{Unsupervised techniques for speech enhancement}
Although supervised techniques of speech enhancement and separation have achieved great success towards improving speech intelligibility, the inherent problems associated with supervised learning still prohibits their applications in all scenarios. First, collecting parallel data of clean and noisy (mixed) data remains costly and time consuming. This limits the amount of data that can be used to train these models. Consequently,  the models are not exposed to enough variations of the recording during training hence affecting their generalizability to  noise types and acoustic conditions that were not seen during training \cite{bie2022unsupervised} \cite{fujimura2021noisy}. Collecting clean audio is always difficult and requires a well-controlled studio exacerbating the already high cost of data annotation \cite{fujimura2021noisy}. Unsupervised learning offers an alternative to solving these problems. The existing unsupervised techniques for speech enhancement and separation can roughly be categorised into three: MixIT based techniques, generative modelling technique  and teacher-student based techniques. Few novel techniques have also been proposed that fall outside these three dominant categories. Work in \cite{Wisdom2020} proposes mixture invariant training (MixIT) to perform unsupervised speech separation. Given a set of $X$ that is composed of mixed speech i.e. $X=\{x_1,x_2,\cdots,x_n\}$ where each mixture $x_i$ is composed of up to $N$ sources, mixtures are drawn at random from the set $X$ without replacement and a mixture of mixture (MoM) created by adding the drawn mixtures, for example if two mixtures $x_1$ and $x_2$ are drawn from the set $X$, $MoM$ $\bar{x}$ is created by adding $x_1$ and $x_2$ i.e $\bar{x}=x_1+x_2$. The MoM $\bar{x}$ is the input to a DNN model which is trained to estimate sources $\hat{s}$ composed in $x_1$ and $x_2$. The DNN model is trained to minimize the loss function in equation 57. 
\begin{equation}
    L_{MixIT}=\min_{A} \sum_{i=1}^{2}L(x_i,[A\hat{s}]_i)
\end{equation}
For a case where MoM is composed of only two mixtures, $A\in B^{2\times M}$ is a set of binary matrices where each column sums to 1. The loss function is trained to minimize the loss between mixtures $x_i$ and the  remixed separated sources $ A\hat{s}$. MixIT has been criticised for over-separation  where it outputs estimated sources greater than the  actual number of underlying sources in the mixtures $x_i$ \cite{zhang2021teacher}. Further, MixIT does not work well for speech enhancement ( i.e., denoising) \cite{saito2021training}. MixIT teacher-student unsupervised model has been proposed in \cite{zhang2021teacher} to  tackle the problem of over-separation in MixIT. It trains a student model such that its output matches the number of  sources  in the mixed speech $x$. Another MixIT based technique for solving over-separation problem is discussed in MixCycle \cite{karamatli2022mixcycle}.  Work in \cite{trinh2022unsupervised} proposes to improve MixIT to make it more tailored for denoising by exploiting an ASR pre-trained model to modify MixIT's loss function. Work in \cite{saito2021training} also seeks to improve MixIT for denoising by improving loss function and  noise augmentation scheme. RemixIT \cite{tzinis2022remixit} is an unsupervised speech denoising tool that exploits teacher-student DNN model. Given a batch of noisy speeches of size $b$, the teacher estimates the clean speech sources $\hat{s}_i$ and noises $\hat{n}_i$ where $1\leq i\leq b$. The teacher estimated noises $\hat{n}_i$ are mixed at random to generate $n^p$. The mixed noise $n^p$ together with the teacher estimated sources are used to generate new mixtures $\hat{m}_i=\hat{s}_i+n^p$. The new mixtures  $\hat{m}_i$ are used as input to the student. The student is optimised to generate $\hat{s}_i$ and noise $n^p$ i.e., $\hat{s}_i$ and $n^p$ are the targets. Through this the teacher-student model is trained to denoise the speech. In RemixIT,  a pre-trained speech enhancement model is used as the teacher model. Motivated by RemixIT, \cite{chentraining} also  proposes a speech denoising unsupervised tool using teacher-student DNN model. They propose various techniques of student training.  MetricGAN-U \cite{fu2022metricgan} is an unsupervised GAN based speech enhancement tool that trains a conditioned GAN discriminator without a reference clean speech,  MetricGAN-U  employs objective in equation 58 to train the speech enhancement model. 
\begin{equation}
 L=\mathbb{E}_{x}[ (D(G(x)) -Q\prime(G(x)))^2+ (D(x)-Q\prime(G(x))^2]
\end{equation}
In equation 58, $Q\prime$ is a non-intrusive metric (i.e does not require reference clean speech) that is used to score the enhanced speech from the generator. The scores obtained by $Q\prime$  are used to optimize the model. In MetricGAN-U, DNSMOS \cite{reddy2021dnsmos} is used as $Q\prime$. Another GAN based technique for unsupervised learning is \cite{xiang2020parallel} which exploits CycleGAN \cite{zhu2017unpaired} multi-objective learning to perform parallel-data-free speech enhancement. Tools in \cite{bie2022unsupervised} and \cite{li2021domain} propose unsupervised speech denoising technique based on variations of VAE. Work in \cite{fujimura2021noisy} proposes a speech denoising technique that uses only the noisy speech. It exploits the idea that was first proposed in \cite{lehtinen2018noise2noise} where they demonstrated that it is possible to recover signals under  corruptions without observing clean signals. Predicating their work on these findings, given a noisy speech signal $x$, and noise $n$, \cite{fujimura2021noisy} creates a more noisy speech $y=x+n$. They then train a DNN model to predict an enhanced speech $\hat{s}$ by having the more noisy input $y$ as the input and noisy speech $x$ as the target. Consequently, the DNN is trained by minimizing the loss in equation 59. 
\begin{equation}
    L=\frac{1}{M}\sum_{i=1}^{M}D(\hat{s}_m,x_m)
\end{equation}
Here, $D$ is the objective function and $M$ is the sample size.
This technique works on the basis that DNN cannot predict random noise hence the noise component in the training data is mapped to their expected values. Therefore, by assuming the noise as zero mean random variable, the objective function  eliminates the noise \cite{fujimura2021noisy}.  Work in \cite{Wang2017b} proposes unsupervised techniques to perform speech separation based on gender. They exploit i-vectors to model large discrepancy in vocal tract, fundamental frequency contour, timing, rhythm, dynamic range, etc between speakers of different genders. In this case DNN model can be viewed as gender separator.

\section{Domain adaptive techniques for speech enhancement and Separation}
 Training data used to train speech enhancement and separation tools mostly have acoustic features that are significantly different from the acoustic features of the speech signals where the tools are deployed. This mismatch between the training data and target data  leads to degradation in the tool’s performance in their deployed environment \cite{pan2010domain}.The target environment dataset's acoustic features may vary from the training data in  noise type, speaker and signal-noise-ratio \cite{li2021domain}. One potential way of tackling this problem is to collect massive training data that covers different variation of deployment environment. However, this is mostly not possible due to prohibitive cost. Due to this, some tools are proposing DNN based techniques for domain adaptation. Domain adaptation seeks to exploit either labelled or unlabelled target domain data to transfer a given tool from training data domain to the target data domain. Basically, domain adaptation seeks to reduce the  covariance shift between the source and target domains data. The domain adaptation techniques in literature for speech separation and enhancement tools can be categorised into two: Unsupervised domain adaptation techniques such as \cite{liao2018noise} \cite{wang2018unsupervised} which use unlabelled target domain dataset to adapt a DNN model and supervised domain adaptation techniques such as \cite{xu2014cross} \cite{li2021domain} \cite{pascual2018language} which exploit limited labelled target domain dataset to perform domain adaptation of a DNN model for speech enhancement or separation. To make speech enhancement tools portable to new a new language, \cite{xu2014cross} proposes to use transfer learning. Transfer learning entails tailoring trained  DNN models to apply  knowledge acquired during training  to a new domain where there is some commonality in type of task. The tool fine-tunes the top layers of a trained DNN model for speech enhancement by using labelled data of a new language while freezing the lower layer which are composed of parameters acquired during training of the original language. Work in \cite{pascual2018language} also uses transfer learning to show that  pre-trained SEGAN can achieve high performance in new languages  with unseen speakers and noise with just short training time. To make it more adaptable to different types of noise, tool in \cite{li2021domain} proposes to employ multiple encoders where each encoder is trained in a supervised manner  to focus only on given acoustic feature. The features are categorized into two i.e., utterance-level features such as gender, age, ascent of the speaker, signal-to-noise ratio and noise type and the signal-level features such high and low frequency of the speech parts. Feature focused encoders are trained to learn how to extract a given feature representation such as gender representation composed in the speech. Through the feature focused encoders, the experimental results show that the tool can adapt more to unseen noise types as compared to using a single global encoder. To adapt the DNN speech enhancement model to unseen noise type, work in \cite{liao2018noise} utilizes domain adversarial training (DAT)\cite{ganin2016domain} to train an encoder to extract noise-invariant features. To do this, it utilizes the labelled source data and unlabelled target data. Through the feedback from the discriminator which gives the probability distribution over multiple noise types, the encoder is trained to produce noise-invariant features, hence reducing the mismatch problem. Work in \cite{wang2018unsupervised} also exploits  unsupervised DAT for speaker mismatch resolution. Work in \cite{li2021domain} exploits importance-weighting (IW) using the classifiers of the networks  to classify the source domain samples from the outlier weights and hence reduces the shift between the source and target domain.
\section{Use of pre-trained models in speech separation and enhancement}
Pre-trained models have become popular especially in Natural language processing(NLP) and Computer vision. In NLP, for example, large corpus of text can be used to learn  universal language representations which are beneficial for downstream NLP tasks. Due to their success in domains such as NLP and computer vision, pre-trained models based on unsupervised learning have been introduced in audio data \cite{Chung2019} \cite{Chung2020} \cite{Liu2020} \cite{Liu2021}\cite{wav2vec200}\cite{Hsu2021}. Such pre-trained models are beneficial in several ways: 
\begin{enumerate}
    \item  Pre-trained models are trained in  large speech dataset hence can learn universal speech representations which can be beneficial  to speech separation by boosting the quality of enhanced speech generated by these models.
    \item Pre-trained models provide models with better initialization which can result in better generalization and speed up convergence during training of speech enhancement models.
    \item Pre-trained speech models can act as regularizers to help speech enhancements models to avoid over fitting.
    \end{enumerate}
Work in \cite{Huang2022} seeks to  establish if  pre-trained speech  models will help generate more robust features for downstream speech denoising and separation task as compared to features established without pre-trained models. To do this they use 13 speech pre-trained models to generate features of a noisy speech which are then passed through a three-layer BLSTM network which generates speech denoising or separation mask. They compare the performance of these features with those of baseline  STFT and  mel filerbank (FBANK) features. Their experiments establish that the 13 pre-trained models used do not significantly improve feature representations as compared to those of baselines. Hence the quality of enhanced and separated speech generated by features of pre-trained models are only slightly better or worse in some cases as compared to those generated based on the baseline features. They attribute this to domain mismatch and information loss. Since most of the pre-trained models were trained with clean speech, they are not portable to a noisy speech domain. Pre-trained models are usually trained to model global features and long-term dependencies hence some local features of the noisy or mixed speech signal may be lost due to this during feature extraction. Using HuBERT Large model \cite{hsu2021hubert}, they demonstrate that the last layer of the model does not produce the best feature representation for speech enhancement and separation. In fact, for speech separation, the higher layers features are of low quality as compared to lower layers. They show that the weighted-sum representations of the representations from the different layers  of pre-trained models where lower layers are given more weight generate better speech the enhancement and separation results as compared to isolated layers representations.  They hypothesise that this could be due  loss of some local signal information necessary for speech reconstruction tasks  in deeper layers. To address the problem of information loss in pre-trained model, \cite{hung2022boosting}  proposes two solutions, first they  utilize cross-domain features as model inputs to compensate lost information and secondly, they fine-tune a pre-trained model by using a speech enhancement model such that the extracted features are more tailored towards  speech enhancement. Research in \cite{irvin2023self} seeks to synthesise clean speech directly from a noisy one using pre-trained model and HiFiGAN \cite{kong2020hifi} speech synthesis model. It exploits the pre-trained model to extract features of the noisy model. The features are then used as input of HiFiGAN which generates estimated clean speech from these features. Based on the results reported in  \cite{Huang2022} that demonstrated that the final layer of pre-trained model does not give optimized representation for speech enhancement, they exploit weighted average of the representations of all the layers of the pre-trained model to generate representations of the noisy speech. The novelty of this work is that they do not use a model dedicated for speech denoising rather show that given features of a noisy speech, speech synthesis model can perform denoising. In \cite{germain2018speech}, a pre-trained model has been exploited to design the loss function. Given a pre-trained model $\Phi$ with $m$ layers, the tool uses weighted $L^1$ loss to compute the difference between the feature activations of clean and noisy speech  generated by different layers of the pre-trained model according to equation 60.
\begin{equation}
    L=\sum_{i=1}^{m}\lambda_w||\Phi_m(s)-\Phi_m(g_\theta(x))||
\end{equation}
Here, $ s$ and $x$ are the clean and noisy speech respectively, $g_\theta$ is the denoising DNN model and $\lambda_m$ are the weights of contribution of each pre-trained model layer features to the loss function.  Work in \cite{hao2023neural} proposes a two-stage speech enhancement framework where they first pre-train a model using unpaired noisy and clean data and utilize the pre-train model to perform speech enhancement. Unlike the previous works that use general pre-trained models for audio, the pre-trained model in \cite{hao2023neural} is trained using speech enhancement dataset. They report state of the art results in speech enhancement and ability of the tool to generalize to unseen noise.

\section{Future direction for research in speech enhancement}

    \textbf{Unsupervised techniques for speech separation}: Majority of speech enhancements tools use supervised learning technique. For those that use unsupervised learning discussed in section 8, they almost entirely focus on speech denoising and not speaker separation and dereverberation. There is therefore a gap of extending  the unsupervised DNN techniques  to perform multi-speaker speech separation and dereverberation.\\
    \textbf{Dimension mismatch problem}: Most speech separation tools set a fixed number $C$ of speakers therefore cannot deal with an inference mixture with $K$ sources, where $C\neq K$.  Currents tools deal with the dimension mismatch problem by either outputting silences if $C>K$ or performing speech separation through iteration. However, both techniques have been found to be inefficient (see discussion in section 2.1). Therefore, there need to explore on developing  DNN techniques for speech separation which are dynamic to the number of  speakers present in inference mixture and adapt appropriately.\\
    \textbf{Focus on data}: Most model compression techniques for speech enhancement speed up the model performance by reducing or optimizing the model parameters. They don't focus on the data-side impact on the model performance. For instance, what is the ideal  sequence length and  chunk overlap when working in time-domain  that can speed up the speech enhancement process without compromising the quality of enhancement. More focus needs to turn towards  exploring the data modifications that can speech up the speech enhancement process.\\
    \textbf{Dataset modification}: In dereverberation, tools are beginning to explore the use of speech with early reverberation \cite{Valin2022} as the target as opposed to using anechoic target. Experiments in \cite{Valin2022} demonstrate that allowing early reverberation in the target speech improves the quality of enhanced speech. There is need to develop a standardized dataset where the target is composed of early reverberation to allow for standardized evaluation of the tools on this dataset.\\
    \textbf{Pre-trained Model}: The pre-trained models that have been utilized for speech enhancement or separation have been trained on clean dataset hence failing potability test when used to generate features of a noisy speech signal \cite{Huang2022}. There is need for development of a pre-trained model tailored for speech separation and enhancement.
 
\section{Conclusion}
This review gives a discussion  on how DNN techniques are being exploited by  monaural speech enhancement tools. The objective was to uncover the key trends and dominant techniques being used by DNN tools at each stage of  speech enhancement process. The review therefore discusses the type of features being exploited by these tools, the modelling of speech contextual information, how the models are trained (both supervised and unsupervised), key objective functions, how pre-trained speech models are being utilized and  dominant dataset for evaluating the performance of the speech enhancement tools. In each section we highlight the standout challenges and how tools are dealing with these challenges. Our target is to give an entry into the speech enhancement domain by getting a thorough  overview of the concepts and the trends of the domain. The hope is that  the review gives a snapshot of the current research on  DNN application to speech enhancement.

\bibliography{template}
 \bibliographystyle{IEEEtranN}
\end{document}